%% file: dataset.tex
\documentclass[%
aps,
prd,
amssymb,
amsmath,
amsfonts,
eqsecnum,
showpacs,
nofootinbib,
floatfix,
twocolumn,
twoside,
a4paper,
superscriptaddress,
longbibliography
]{revtex4-2}

\input{UsePackages}
\usepackage{graphicx}% Include figure files
\usepackage{dcolumn}% Align table columns on decimal point
\usepackage{bm}% bold math
\usepackage{tabularx}
\usepackage{multirow}
\usepackage{capt-of}
\usepackage{color}
\usepackage{url}
\usepackage[utf8]{inputenc}
\usepackage{placeins}
\usepackage{comment}

\usepackage{natbib}

\usepackage[nolist,nohyperlinks]{acronym}
\usepackage{xspace}
\usepackage[colorlinks]{hyperref}

\usepackage[caption=false]{subfig}

\DeclareMathAlphabet{\mathbfsf}{\encodingdefault}{\sfdefault}{bx}{sl}

\newcommand{\be}{\begin{equation}}
\newcommand{\ee}{\end{equation}}
\newcommand{\bea}{\begin{eqnarray}}
\newcommand{\eea}{\end{eqnarray}}

\newcommand{\Shat}{\hat{S}}

\newcommand{\phX}{\texttt{IMRPhenomX}\xspace}

\newcommand{\phT}{\texttt{IMRPhenomT}\xspace}

\newcommand{\NRSur}{\textsc{NRSur7dq4}\xspace}
\newcommand{\NRSurE}{\textsc{NRSur7dq4EmriRemnant}\xspace}

\newcommand{\cone}{\ensuremath{\chi_1}}
\newcommand{\ctwo}{\ensuremath{\chi_2}}

\definecolor{dodgerblue}{HTML}{1E90FF}
\definecolor{viennared}{HTML}{DA0A14}
\definecolor{ctorange}{HTML}{FF6C0C}
\definecolor{granadagreen}{HTML}{078931}
\definecolor{wales}{HTML}{ff0038}
\definecolor{valenciacfred}{HTML}{ee3524}
\definecolor{barcelonafcgold}{HTML}{edbb00}
\definecolor{jam}{HTML}{A50B5E}
\definecolor{austriawien}{HTML}{441678}

\AtBeginDocument{%
  \hypersetup{
    citecolor=dodgerblue,
    linkcolor=dodgerblue,   
    urlcolor=dodgerblue}}

\newcommand{\UIB}{Departament de F\'isica, Universitat de les Illes Balears, IAC3 -- IEEC, Crta. Valldemossa km 7.5, E-07122 Palma, Spain}

\newcommand{\ICE}
{Institut de Ci\`encies de l'Espai (ICE, CSIC), Campus UAB, Carrer de Can Magrans s/n, 08193 Cerdanyola del Vall\`es, Spain}

\allowdisplaybreaks[1]

\begin{document}

% ~~~~~~~~~~ Title & Abstract ~~~~~~~~~~ %
\title[ML]
{Building a bridge between comparable and extreme mass ratio black hole binaries: a single spin precessing model for the final state}

\author{Maria de Lluc Planas}
\affiliation{\UIB}

\author{Joan Llobera-Querol}
\affiliation{\UIB}

\author{Sascha Husa}
\affiliation{\ICE}
\affiliation{\UIB}

\date{\today}

\begin{abstract}
Modelling the gravitational wave signal from binaries beyond comparable mass is an important open issue in gravitational wave astronomy. For non-spinning binaries and when the spins are aligned with the orbital angular momentum, some first studies concerning the transition between the comparable and extreme mass ratio regime are already available, which suggest that extreme mass ratio results at times extrapolate to comparable mass ratios with surprising precision. Here we study the case of misaligned spins: We present new numerical relativity (NR) simulations performed with the Einstein Toolkit code at mass ratios up to 18 and construct a heterogeneous dataset that spans all mass ratios, including data from NR simulations, numerical approximations to extreme mass ratio binaries, and data from the geodesic approximation. As a first application we provide fits for the remnant mass and spin magnitude in single spin precessing systems, omitting consideration of the in-plane spin orientation. These fits demonstrate accuracy comparable to the state-of-the-art \NRSurE model, all while retaining the simplicity and efficiency inherent in previous phenomenological fits.
\end{abstract}


\maketitle

%%%%%%%%%%%%%%%%%%%%%%%%%%%%%%%%%%%%%%%%%%%%%%%%%%%%%%%%%%%%%%%%%
%               INTRODUCTION
%%%%%%%%%%%%%%%%%%%%%%%%%%%%%%%%%%%%%%%%%%%%%%%%%%%%%%%%%%%%%%%%%
\section{Introduction}
\label{sec:Introduction}
The LISA space mission is expected to observe gravitational waves from compact binaries with a large range of mass ratios \cite{AstroLISA, eLISA} up to and including extreme mass ratio inspirals (EMRI). EMRI waveforms are best described via the self-force method, where one perturbs in the mass ratio of the system (\cite{selfforce1, selfforce2}, see \cite{selfforceReview} for discussion). 
For comparable mass binaries, several families of waveform models have been developed \cite{IMRPhenomTPHM, IMRPhenomXPHM, IMRPhenomXODE, SEOBNRv4PHM, SEOBNRv5PHM, TEOBResumS, NRSur7dq4, NRHybSur3dq8}, which have become indispensable tools for gravitational wave data analysis, e.g.
\cite{LIGOScientific:2018mvr,LIGOScientific:2020ibl,LIGOScientific:2021usb,LIGOScientific:2021djp}.
Such waveform models are calibrated to data from numerical relativity (NR) simulations \cite{PhenomPNR, PhenomXO4a, PhenomXO4abis}. However, for the foreseeable future, NR waveforms will be sparse in the parameter space of precessing (even more so for generic) black hole binaries, especially at high mass ratios (see \cite{RIT4, SXS2, Cardiff-catalog, MAYA2, CoRe2} for the latest releases of NR catalogs). While simulations at, say, mass ratios of order $10^2$ or $10^3$ are in principle feasible in numerical relativity \cite{Lousto:2022hoq}, the number, length, and accuracy of such waveforms will be severely restricted by computational cost until new computational approaches are developed (see however \cite{Fernando:2022php}). 
Since the black hole binary population is not yet well understood, it will be prudent to develop and deploy waveform models that accurately describe a wide range of mass ratios well before the LISA era. 

It is well known that the extreme mass ratio limit 
and the self force expansion in mass ratio can provide useful information even for the comparable mass ratio regime \cite{LeTiec2013,Albertini:2022rfe,Wardell:2021fyy}, fueling hope that using such information can reduce the number of NR waveforms that are needed for calibrating waveform models.
To date, work that connects the two mass ratio regimes has focused on either non-spinning binaries \cite{BHPTNRSur1dq1e4, BHPTNR_Remnant, Interplay}, 
the use of extreme mass ratio waveforms to calibrate quasi-circular aligned spin waveform models \cite{SEOBNRv5PHM, TEOBResumS, phenomxhm, phenomthm}, or the simpler problem of models for the final mass and spin in the quasi-circular aligned spin case \cite{Jimenez-Forteza:2016oae}. For misaligned spins one however faces much more complicated phenomena, and a much larger parameter space (7 instead of 3 dimensions in the absence of orbital eccentricity, as is the case here).

In this work, we make a first step to bridge the comparable mass and extreme mass ratio regimes in the misaligned spin sector, and develop precessing models for the remnant mass and spin, thus extending previous work \cite{Jimenez-Forteza:2016oae} to the precessing case. In order to simplify the analysis and focus on the transition to large mass ratios we restrict the models to the single spin case, and leave  double spin effects for future work. %\cite{IMRPhenomXPHM, PhenomPNR}.
As our input data we construct a consistent heterogeneous dataset for quasi-circular precessing binaries, which combines NR waveforms from different codes, numerical solutions of the Teukolsky equation \cite{PhysRevD.100.084031, PhysRevD.100.084032}, and information from Kerr geodesics \cite{Schmidt_2002}. 
To understand the region where no NR information is available, we rely on approximations based on Kerr geodesics, which surprisingly provide valuable information across the parameter space, i.e. from EMRIs to comparable mass binaries.
Recent remnant models in the literature include aligned spin datasets with and without information from the extreme mass ratio (EMR) limit, see for instance Refs.~\cite{Healy2014, Jimenez-Forteza:2016oae}, precessing datasets with information from the aligned spin EMR limit, see \cite{PhysRevD.92.024022,EMRIfinalspin}, surrogate models like \NRSurE \cite{NRSur7dq4EmriRemnant}, and also a machine learning approach \cite{Haegel:2019uop}.

In Sec.~\ref{sec:construction} we describe the datasets we use, and the procedures chosen to blend them into a single consistent dataset.
In the quasi-circular aligned spin case, generating an heterogeneous dataset is relatively straightforward, since
the intrinsic parameters only consist of the masses and spins, which in turn depend only very weakly on time due to the very small amounts of infalling radiation. Hence, the time dependence of masses and spins is often neglected in aligned spin waveform models \cite{phenomxhm, phenomthm}. 
For misaligned spins however, the spin angles and orientation of the orbital plane depend on time, and a coordinate frame needs to be defined judiciously to consistently parameterize the different datasets.

In Sec.~\ref{sec:applications}
we use our heterogeneous dataset to compute the remnant mass and spin magnitude across all mass ratios for the case when only the larger black hole is spinning, as the spin on the smaller black hole becomes a subdominant effect for large mass ratios~\cite{Mathews:2021rod}. 
Additionally, we neglect the orientation of the in-plane component of the single spin, a decision driven by the current limitations in computational cost and tests of the impact of the in-plane angle on our results.
Ultimately, a careful selection of the quantities for modeling allows us to generate simple parameterized fits for both the mass and spin magnitude of the remnant in a precessing binary. These fits can be evaluated efficiently and achieve an accuracy comparable to the \NRSurE model \cite{NRSur7dq4EmriRemnant}, the current state-of-the art in remnant models.

Finally, in Sec.~\ref{sec:conclusions}, we summarize and discuss the scope and limitations of the work we report here, as well as next steps.

Throughout this paper we use geometric units with $G=c=1$. Component masses are denoted by $m_i$, we define the mass ratio $q = m_1/m_2 \geq 1$, and the symmetric mass ratio
$\eta = {m_1 m_2}/{(m_1+m_2)^2}$. The total component mass will be denoted by $M = m_1 + m_2$ and will serve as a scale parameter.
The dimensionless spin magnitudes are denoted~$\chi_i$. 

\section{Precessing dataset}\label{sec:construction}
%%%%%%%%%%%%%%%%%%%%%%%%%%%%%%%%%%%%%%%%%%%%%%%%%%%%%

\begin{figure} 
    \includegraphics[width=1\columnwidth]{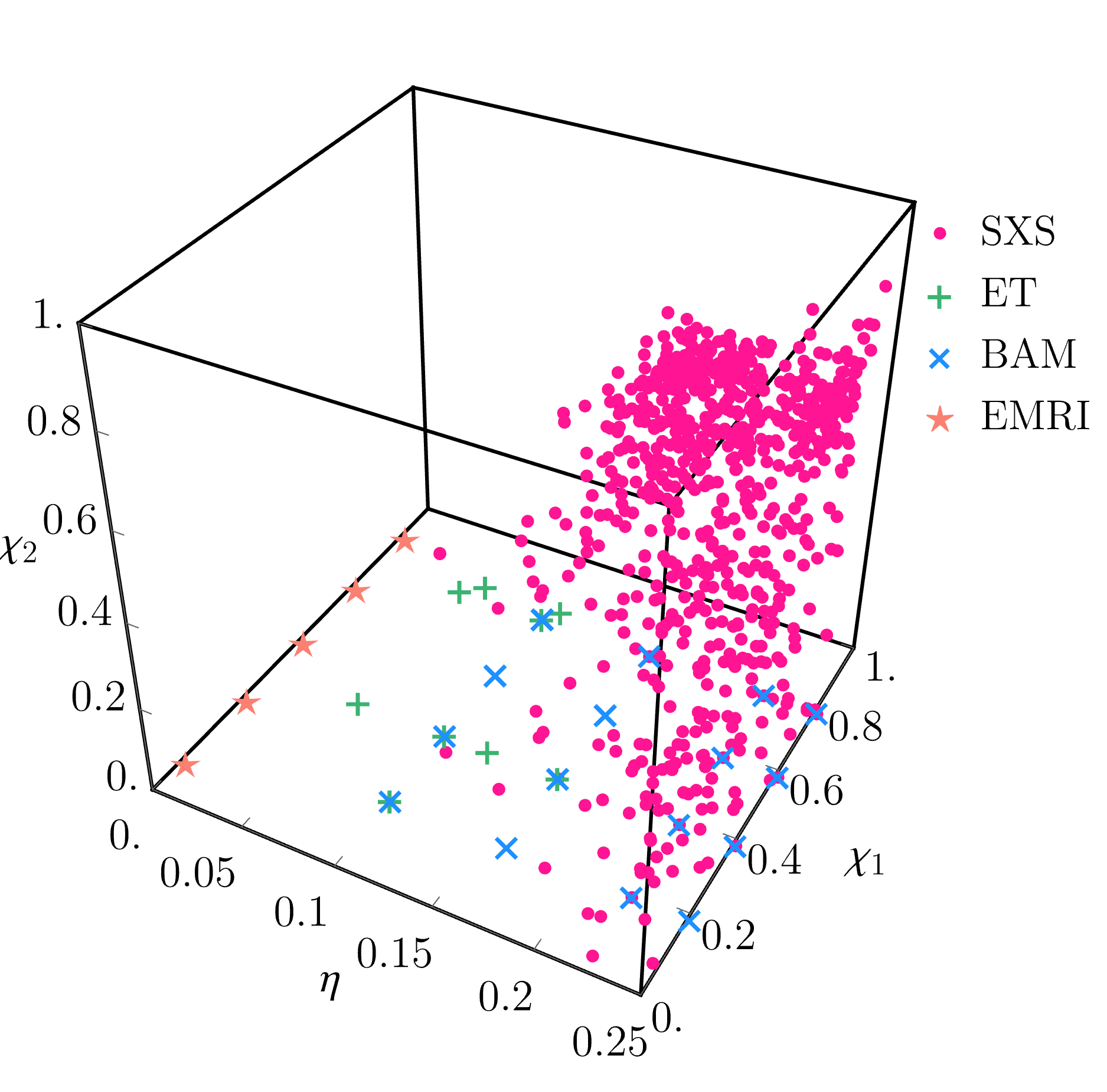}
    \caption{Three-dimensional representation of the precessing numerical relativity simulations used in this project described in Sec.~\ref{sec:construction}. The visualization presents the distribution of the data in terms of their symmetric mass ratio $\eta$ and the spin magnitudes of the largest and smallest black holes, denoted as $\chi_1$ and $\chi_2$ respectively.
    \label{fig:dataset}}
\end{figure}

\begin{figure} 
    \includegraphics[width=1\columnwidth]{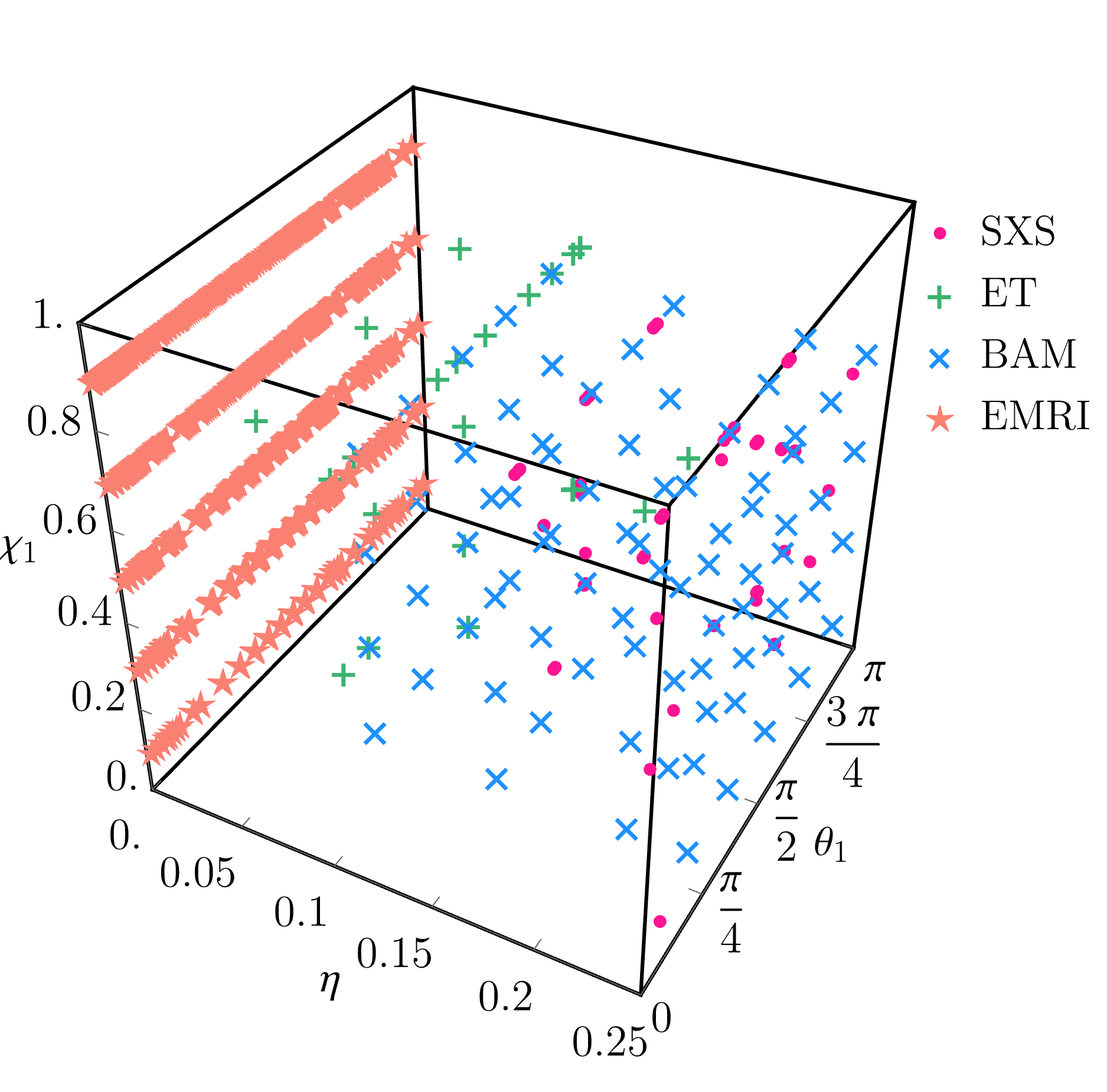}
    \caption{
    Three-dimensional representation of the single spin subset of the precessing numerical relativity simulations introduced in Sec.~\ref{sec:construction}. It shows the distribution of the data in terms of their symmetric mass ratio $\eta$, the spin magnitude of the largest black hole $\chi_1$, and its orientation with respect to the orbital frequency vector at the reference time $\theta_1$. 
    \label{fig:datasetSS}}
\end{figure}

In Sec.~\ref{sec:NR}-\ref{sec:EMRIs} we review the ingredients for our heterogeneous dataset:
\begin{itemize}
\item NR waveforms from the public SXS catalogue \cite{SXS2}, produced with the SpEC code \cite{SpEC}, and the Cardiff group's public catalogue \cite{Cardiff-catalog} of waveforms produced with the BAM \cite{BAMcode, 6thorder} code.
\item NR waveforms recently produced with the public Einstein Toolkit code \cite{ETK} which have not been presented previously.
\item Numerical solutions of the Teukolsky equation for inspirals at mass ratio $1000$ ~\cite{PhysRevD.100.084031,PhysRevD.100.084032}.
\item Solutions of the geodesic equation in Kerr spacetime.
\end{itemize}
In Sec.~\ref{sec:blending} we discuss how to blend all the above data into one consistent dataset for the remnant mass and spin.

Our datasets use different coordinate gauge conditions, and thus spin angles can not be expected to be exactly the same even for physically identical binaries. We expect such uncertainties to only correspond to a few degrees  \cite{Ossokine2015}, and to not play a major role at our current level of accuracy. In our work we again find approximate consistency between different datasets, this aspect will however have to be studied further in the future.

\subsection{NR datasets}
\label{sec:NR}

Our NR dataset spans mass ratios from $q=1$ (equal masses) to $q=18$. We use data from two publicly available catalogues of NR simulations, the SXS catalog \cite{SXS2}, obtained with the SpEC code \cite{SpEC}, and a catalog of waveforms \cite{Cardiff-catalog} obtained with the BAM code \cite{BAMcode, 6thorder}. In addition we use new simulations we performed with the Einstein Toolkit (ETK) \cite{ETK}.
The SXS simulations are performed with the generalized harmonic formulation of the Einstein equations \cite{Pretorius}, while the BAM and ETK simulations use the moving puncture setup with the 1+log lapse 
and $\tilde\Gamma$-driver shift coordinate conditions. 
We have analyzed both the waveforms and apparent horizon data of all simulations to create a consistent heterogeneous dataset encompassing information from both sources. In this paper we however only discuss the remnant properties, leaving investigations into the precessing waveform dataset to future work.

All the NR points in our dataset are included in Fig.~\ref{fig:dataset}, where we show the distribution of the data in a three-dimensional subspace defined at $100\ M$ before merger.
The merger time is not defined in exactly the same way for data produced with different codes. We report details for each catalog in the subsections below, however our findings indicate that the small differences in the definition of the merger time across catalogs does not significantly impact the results at the current level of accuracy. For all datasets we shift the time coordinate to the value of zero at the merger time.
From Fig.~\ref{fig:dataset} we can see that the majority of the points are concentrated in the comparable mass regime ($\eta \leq 0.15$), mostly from the SXS catalogue. The BAM points are located in the lower plane as they are single spin simulations, while the high mass ratio ETK simulations are dispersed in the mid-high mass ratio regime.
In Fig.~\ref{fig:datasetSS} we show the single spin simulations distributed in the $(\eta,\chi_1,\theta_1)$-parameter space, where $\chi_1$ is the magnitude of spin of the more massive black hole, and $\theta_1$ the angle between the spin and the axis of orbital motion $100 M$ before merger. These are the simulations we are using to compute the remnant fits in Sec.~\ref{sec:applications}.

\subsubsection{SXS}\label{sec:sxs}
%%%%%%%%%%%%%%%%%%%%%%%%%%%%%%%%%%%%
We use 1409 quasi-circular precessing simulations from the SXS catalogue \cite{SXS2}, which range from mass ratio $1$ to $6$ and $0<\chi_1<0.99$, $0<\chi_2<0.9$. To confine the parameter space to quasi-circular orbits, we impose a limit on the orbital eccentricity $e$ (effectively defined as the Newtonian eccentricity, see \cite{SXS2}) to $e\leq 0.002$, leading to the exclusion of 13 simulations from the original catalog. 
The \texttt{reference\_eccentricity} parameter from the metadata was utilized for this purpose. 
Among the 1409 simulations included in our analysis, 80 correspond to single spin configurations. The criterion for identifying single spin cases was $\chi_2(t_{\mathrm{ref}})<0.001$.
These specific waveforms are employed for the remnant properties fits presented in this paper, with the reference time set at $t_{\mathrm{ref}}=-100\ M$ (where the merger corresponds to $t=0$).
For the SXS dataset the merger time is defined as the maximum of the $L^2$ norm of all the available modes, as provided by the \texttt{sxs} python package.

The simulations have been performed with the pseudo-spectral SpEC code \cite{SpEC}, which excises spatial regions inside of pure outflow boundaries located inside but close to the apparent horizons of the black holes.
Initial data are constructed to satisfy the constraints of general relativity using the Extended Conformal Thin Sandwich \cite{York:1998hy,Lovelace:2008tw} equations.

\subsubsection{BAM dataset}\label{sec:BAM}
%%%%%%%%%%%%%%%%%%%%%%%%%%%%%%%%%%%%%%%%%%%%

We also use 80 simulations from the single spin Cardiff precessing catalog \cite{Cardiff-catalog}, which span the $1\leq q \leq 8$, $0<\chi_1<0.8$, $\chi_2=0$ parameter space evenly in mass ratio, spin magnitude and orientation $\theta_1$, so $\theta_1(t_{\mathrm{in}})\in \left\{\frac{\pi }{6},\frac{\pi }{3},\frac{\pi }{2},\frac{2 \pi }{3},\frac{5 \pi }{6}\right\}$. The in-plane orientation of the single spin $\phi_1$ was chosen to be 0 at the relaxed time for each simulation.
The eccentricity for all the simulations was reduced to $0.002$ through manual iterations of the linear momenta of the punctures in the initial parameters (see  Sec. II B 1 in~\cite{Cardiff-catalog} for details).
For the BAM dataset the merger time is defined to be the peak of the $l=2$ multipole modes of $\Psi_4$, provided in the metadata.

The simulations have been carried out in the ``moving puncture'' approach with the BAM code, which uses conformally flat Bowen-York puncture initial data \cite{BowenYork}. Note that this setup for the initial data allows to analytically compute the initial orbital angular momentum analytically using the Newtonian physics formula as a cross product of position vector and linear momentum.
The initial data are evolved with fixed mesh-refinement and sixth order finite differencing \cite{BAMcode, 6thorder}.

\subsubsection{Einstein Toolkit dataset}\label{sec:ET}
%%%%%%%%%%%%%%%%%%%%%%%%%%%%%%%%%%%%%%%%%%%%%%%%%%%%%%%%%%%%

In addition to the public data from the SXS and BAM dataset we also use higher mass ratio simulations we have recently performed with ETK \cite{ETK}. 
We produced 24 simulations which span the region $4\leq q \leq 18$, $0.4 \leq \chi_1 \leq 0.8$, $0 \leq \chi_2 \leq 0.4$.
Due to an inappropriate configuration of the wave extraction grids, some gravitational wave signals exhibit excessive noise, making it challenging to extract the merger time directly from it. We thus rely on horizon quantities, identifying the merger time as the transition from the individual black hole spins to the remnant spin.
Detailed information on these simulations can be found in Table~\ref{tab:ET}.

The setup of our Einstein Toolkit code is very similar to that of the BAM code. Differences include the use of 8th order accurate finite difference stencils,
and the eccentricity reduction algorithm described in \cite{eccredToni}.
Furthermore, fixed mesh refinement with moving cubical boxes is not used for the whole computational domain. However, for the wave extraction region and beyond the multipatch Llama code \cite{LlamaPollney} is used to allow a high radial grid resolution with a reduced memory consumption.
The final mass and spin are determined from the apparent horizons, which are located with the \texttt{AHFinderDirect}-code \cite{Thornburg:2003sf}.

\begin{table*}
\centering
\begin{tabular}{m{0.8cm}m{0.8cm}m{0.8cm}p{2.3cm}m{0.8cm}m{2.1cm}m{2.3cm}m{1.3cm}m{1.3cm}m{2.3cm}m{1.2cm}}
\hline
  ID & $q^{\mathrm{ref}}$ &  $\chi_1$ &  $\theta_{\mathrm{L\chi_1}}$ ( $^{\circ}$ )&  $\chi_2$  &  $\theta_{\mathrm{L\chi_2}}$ ( $^{\circ}$ ) &  $M \omega_{\mathrm{orb}}\left(\times 10^{-2}\right)$ &  $M_f/M$ & $\chi_f$ & $\theta_{L\chi_f}$( $^{\circ}$ ) &  $t_{\mathrm{M}}/M$\\
\hline \hline
   1 & 4 & 0.4 & 135.0 (134.1)  & 0 & - & 1.66 (5.44) & 0.981 & 0.368 & 17.27 (27.89) & 2562 \\ 
   2 & 4 & 0.4 & 170.0 (170.0) & 0 & - & 1.66 (5.29) & 0.982 & 0.265 & 6.41 (9.51) & 2395  \\ 
   3 & 4 & 0.4 & 170.0 (169.8) & 0.4 & 170.0 (170.1) & 1.64 (5.18) & 0.982 & 0.257 & 7.20 (10.14) & 2328  \\
   4 & 6 & 0.4 & 135.0 (134.1) & 0 & - & 1.98 (5.35) & 0.987 & 0.281 & 36.34 (46.30)  & 2555 \\ 
   5 & 6 & 0.4 & 135.0 (133.5) & 0.4 & 135.0 (145.1) & 1.96 (5.31) & 0.987 & 0.283 & 37.67 (47.30) & 2504 \\
   6 & 8 & 0.8 & 170.0 (169.5) & 0 & - & 1.86 (4.83) & 0.992 & 0.285 & 155.1 (156.7) & 2016  \\
   7 & 8 & 0.4 & 90.00 (88.24) & 0 & - & 2.06 (6.32) & 0.989 & 0.440 & 36.77 (43.76)  & 2253 \\ 
   8 & 8 & 0.8 & 135.0 (135.1) & 0 & - & 1.89 (5.21) & 0.991 & 0.460 & 92.79 (100.7)  & 2433 \\ 
   9 & 8 & 0.8 & 90.00 (86.03)& 0 & - & 2.13 (6.54) & 0.986 & 0.700 & 58.90 (59.14) & 2362 \\
   10 & 8 & 0.8 & 170.0 (169.9) & 0 & - & 1.92 (4.80) & 0.992 & 0.285 & 155.0 (157.1) & 2016 \\
   %new isa sims
   11 & 8 & 0.4 & 15.00 (14.89) & 0 & - & 2.49 (7.17) & 0.986 & 0.585 & 6.81 (7.67) & 1605 \\
   12 & 8 & 0.2 & 90.00 (89.98) & 0 & - & 2.63 (6.07) & 0.989 & 0.345 & 24.25 (27.09) & 1047 \\
   13 & 8 & 0.4 & 30.00 (29.14) & 0 & - & 2.49 (7.05) & 0.986 & 0.573 & 13.45 (14.69) & 1574 \\
   14 & 8 & 0.8 & 165.0 (164.5) & 0 & - & 2.19 (4.70) & 0.992 & 0.301 & 144.9 (146.3) & 1307 \\
   15 & 8 & 0.8 & 15.00 (15.13) & 0 & - & 2.61 (9.05) & 0.977 & 0.858 & 9.64 (10.61) & 1906 \\
   16 & 8 & 0.8 & 150.0 (149.2) & 0 & - & 2.24 (4.84) & 0.992 & 0.372 & 117.6 (120.5) & 1274 \\
   17 & 8 & 0.8 & 30.00 (27.55) & 0 & - & 2.58 (8.72) & 0.978 & 0.847 & 19.33 (18.14) & 1895 \\
   18 & 8 & 0.8 & 75.00 (74.63) & 0 & - & 2.62 (6.87) & 0.984 & 0.757 & 49.34 (52.04) & 1368 \\
   19 & 8 & 0.8 & 105.0 (104.1) & 0 & - & 2.47 (5.81) & 0.989 & 0.633 & 68.37 (74.24) & 1320 \\
   20 & 18 & 0.4 & 150.0 (149.9) & 0 & - & 2.51 (5.64) & 0.996 & 0.225 & 124.5 (126.8) & 2234 \\
   21 & 18 & 0.4 & 90.00 (89.34) & 0 & - & 2.74 (6.63) & 0.996 & 0.395 & 61.30 (65.03) & 2221 \\
   22 & 18 & 0.8 & 150.0 (149.4) & 0 & - & 2.39 (5.00) & 0.996 & 0.565 & 138.7 (139.8) & 2207 \\
   23 & 18 & 0.8 & 30.00 (29.78) & 0 & - & 2.93 (10.5) & 0.991 & 0.828 & 24.33 (25.20) & 3243 \\
   24 & 18 & 0.8 & 90.00 (90.35) & 0 & - & 2.87 (6.93) & 0.995 & 0.738 & 75.84 (77.67) & 2243 \\
 \hline
\end{tabular}
\caption{Initial and reference data (in brackets) of the numerical relativity simulations computed using the Einstein Toolkit, described in Sec.~\ref{sec:ET}. The reference time is chosen to be $100\ M$ before merger. From left to right, the table gives the ID of each simulation, its mass ratio, the dimensionless spin magnitude of the larger black hole, the primary spin's orientation with respect to the orbital frequency $\vec{\omega}$, the dimensionless spin of the lighter black hole, its orientation, the dimensionless orbital frequency at which the quantities are given, the dimensionless final mass of the remnant object, its spin, its orientation with respect to $\vec{\omega}$ and finally, the dimensionless merger time.}
\label{tab:ET}
\end{table*}

\subsection{Extreme mass ratio limit}
\label{sec:EMRIs}

In the EMR case ($\eta \ll 1/4 $) one can rely on black hole perturbation theory (BHPT), which assumes a perturbation of Kerr spacetime due to a small object $m_2$ orbiting a black hole of mass $m_1 \gg m_2$.
In the test mass limit ($\eta \rightarrow 0 $) the calculation of the orbital motion decouples from the calculation of the gravitational wave signal, radiation reaction vanishes and the smaller object follows a geodesic. 
Below we first summarize quasi-circular geodesics of Kerr spacetime and then describe our numerical dataset for mass ratio $10^3$ \cite{PhysRevD.100.084031}, which we use for cross-checks.

\subsubsection{Kerr geodesics}
%%%%%%%%%%%%%%%%%%%%%%%%%%%%%%

We consider the geodesic motion of a test mass $m_2$ in a Kerr spacetime of mass $m_1$ and angular momentum $\vert J \vert = a m_1$. For a given black hole with parameters $a$ and $m_1$ the geodesics can be parameterized by the constant orbital quantities 
$p$ (semilatus rectum), $e$ (eccentricity) and $\theta_-$ (inclination parameter), or by the  energy $E$, angular momentum along the axis of symmetry $L_\mathrm{z}$ and Carter's constant $Q$, which are also constants of motion. 
The orbital quantities are defined in Boyer-Lindquist coordinates $(t,r,\theta,\varphi)$ \cite{Boyer:1966qh}.
In this paper we will only be interested in the circular case, where $e = 0$ and radial separation $r$ is constant, while the angular position $\theta$ will be time dependent and oscillates between extrema determined by the inclination parameter $\theta_-$,
\begin{equation}
\theta_{-}\leq\theta\leq ( \pi-\theta_- ). 
\end{equation}

The solution for the time dependent geodesic motion, and the relation between the conserved quantities $(E, L_z, Q)$ and the orbital motion can be found, for instance, in Ref.~\cite{Schmidt_2002}, and is briefly summarized in App.~\ref{app:geodesics}. 
Here we follow the parameterization of \cite{Schmidt_2002} or \cite{Drasco:2005kz}, which is also used in the black hole perturbation toolkit software package~\cite{BHPToolkit}. This toolkit, among other things, provides a Mathematica implementation of Kerr geodesics, \texttt{KerrGeodesics}, which we have used in parallel to our own implementation.
In order to solve for the geodesics and discuss the results it is natural to adopt a scale-invariant formulation of the problem introducing dimensionless quantities:
\begin{equation}\label{eq:dimless}
\tilde{a}=\frac{a}{m_1}\; ,\; \tilde{E}=\frac{E}{m_2}\; ,\; \tilde{L_{\mathrm{z}}}=\frac{L_{\mathrm{z}}}{m_1 m_2}\; , \; \tilde{Q}=\frac{Q}{{m_1}^2 {m_2}^2}.
\end{equation}

The conserved quantities $(\tilde E, \tilde L_z, \tilde Q)$ can be computed algebraically from the parameters $(a, p, e, \theta_-)$, e.g.
\begin{equation}
    \tilde{Q}=\cos^2\theta_{-}\left[\tilde{a}^2(1-\tilde{E}^2)+\frac{\tilde{L_{\mathrm{z}}^2}}{1-\cos^2 \theta_{-}}\right].
\end{equation}
For a discussion of an approximate interpretation of $Q$ as describing the square of the total angular momentum of the particle orthogonal to the 
axis of the black hole see e.g.~\cite{Drasco:2005kz}. There, an equivalent inclination angle $I$ (called $\theta_{inc}$ in ~\cite{Drasco:2005kz}) is defined as
\begin{equation}
I = \frac{\pi}{2} - \mbox{sign}\left(L_z\right)\, \theta_{-},    
\end{equation}
which resembles an alternative definition for an orbital inclination angle $\iota$:
\begin{equation}
\cos \iota = \frac{L_z}{\sqrt{L_z^2+ Q}},    
\end{equation}
where $Q$  plays the role of the magnitude squared of the angular momentum orthogonal to $L_z$. It has been found that in general $\iota \approx I$ and that the angles $I$ and $\iota$ automatically encode a notion of
prograde and retrograde orbits ($I,\iota < 90^\circ$ for prograde and  $I,\iota > 90^\circ$ for retrograde)~\cite{Schmidt_2002}.
In terms of the energy and angular momentum one finds that $\tilde{E}^{(p)}<\tilde{E}^{(r)}$ and $\tilde{L_{\mathrm{z}}}^{(p)}<\tilde{L_{\mathrm{z}}}^{(r)}$, where $p$ stands for prograde orbits and $r$ for retrograde, i.e.  for prograde orbits the particle has higher binding energy.

Of special interest is the innermost stable circular orbit (ISCO). Particles with small but finite mass will adiabatically inspiral to the ISCO, and then plunge into the black hole.
The radiation of energy and angular momentum during the plunge is much smaller than during the inspiral, and the remnant mass and spin can therefore be approximated by the values of the energy and angular momentum at the ISCO. This aspect will be discussed further in Sec.~\ref{sec:applications}, and concretely motivated in Fig.~\ref{fig:comparisongeo}.

The simpler subset of aligned spin binaries is defined by setting the inclination angles $I$ or $\iota$ to $0$ or $\pi$.
The $L_z$ component of the orbital angular momentum then corresponds to the total orbital angular momentum,  the final spin only has a non-vanishing $z$-component, and the orbital plane is preserved.

\subsubsection{Numerical EMRI data}
%%%%%%%%%%%%%%%%%%%%%%%%%%%%%%%%%%%

\begin{figure} 
\includegraphics[width=0.92\columnwidth]{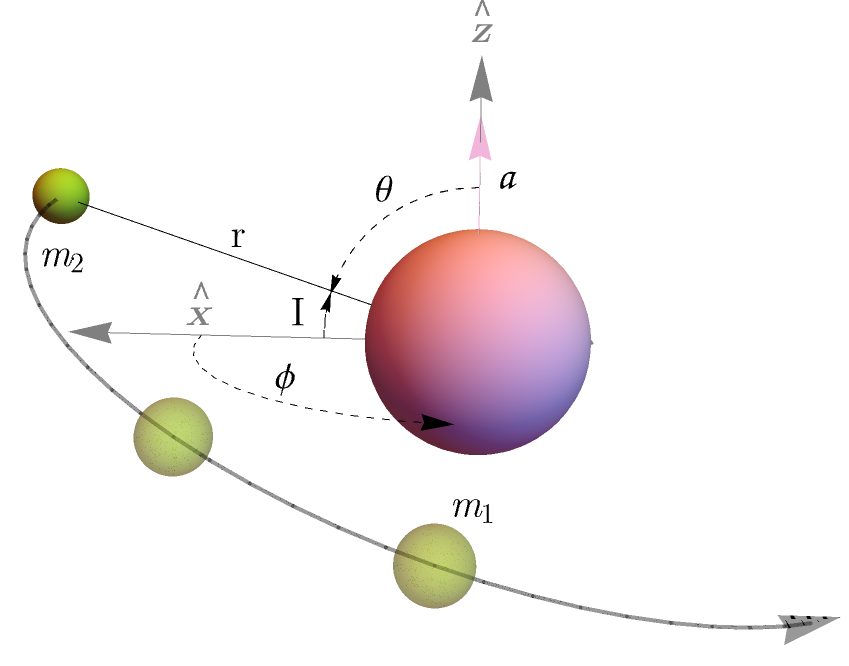}
    \caption{Definition of the EMRI orbital quantities provided in the dataset described in Sec.~\ref{sec:EMRIs}. The dataset spans a parameter space described by the black hole spin magnitude $a$, the inclination angle $I$ and the final angle of the plunge $\theta_f$. The spherical coordinates ($r,\theta,\phi)$ determine the evolved position of the smaller black hole of mass $m_2$.
    \label{fig:EMRI}}
\end{figure}

In this work we use the data of ~\cite{PhysRevD.100.084031},
5925 EMRIs of mass ratio $q=1000$, distributed in a grid of values of the spin of the largest black hole $a$, inclination angle $I$, and plunge angle $\theta_f$ (see Fig.~\ref{fig:EMRI}). Reference~\cite{PhysRevD.100.084031} extends the work of Ori and Thorne \cite{OriThorne} from equatorial to inclined orbits.
The procedure splits the worldline into three regions: i) The adiabatic inspiral, where they use a frequency-domain BHPT code \cite{Drasco:2005kz} to evolve the orbital quantities until they reach ii) the transition region. Closer to the ISCO, the inspiral is no longer adiabatic, requiring further considerations. 
iii) On reaching the plunge, $E,\ L_z$ and $Q$ are frozen to the last value of the orbit while other orbital quantities are evolved solving the geodesic equation. 
We thus define the merger time when the small object crosses the horizon and $E,\ L_z$ and $Q$ reach their frozen value.

In their companion paper~\cite{PhysRevD.100.084032}, they investigate the dependence of individual-mode excitation on plunge parameters via their waveform set. 
The gravitational waves generated by the system are computed from the worldline by solving the Teukolsky equation in the time domain~\cite{PhysRevD.76.104005, PhysRevD.78.024022}. 
Further details on the procedure can be found in Ref.~\cite{PhysRevD.100.084031}.

\begin{figure}   \includegraphics[width=1\columnwidth]{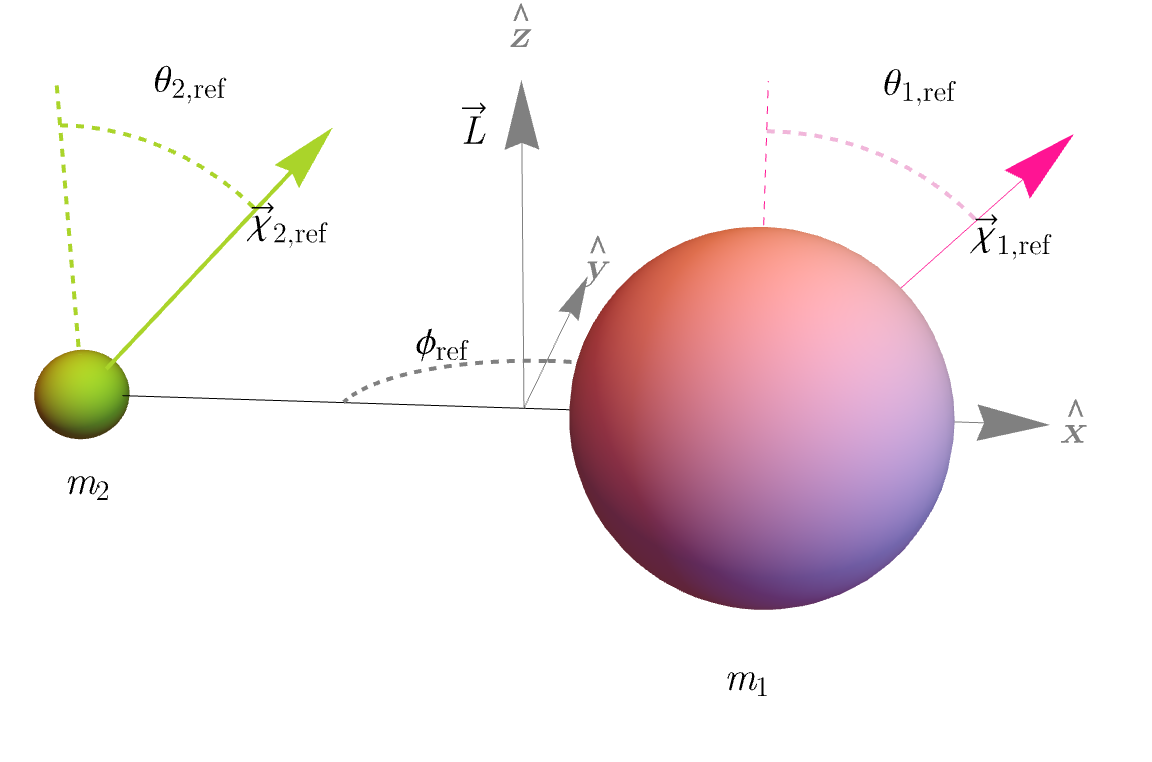}
    \caption{Definition of the orbital quantities employed in this project (traditional to the comparable mass simulations). All vectors are defined with respect to the orbital angular momentum $\vec{L}$, as opposed to those shown in Fig.~\ref{fig:EMRI}, which are specified in terms of the larger black hole spin. Note that in this project we approximate $\vec{L}$ by $\vec{\omega}$~\eqref{eq:omega}.
    \label{fig:coorbital}}
\end{figure}

For each configuration, the dataset includes all the spherical harmonic modes up to $l=4$, plus selected higher $l,m$ modes, and complete information on the small body's trajectory 
\begin{equation}\label{eq:x2}
    \vec{x_2}(t) = -\vec{r} = r(t)(\sin{\theta(t)}\cos{\phi(t)},\sin{\theta(t)}\sin{\theta(t)},\cos{\theta(t)}),
\end{equation}
by providing
($r$, $\phi$, $\theta$), velocity ($\mathrm{d}r/\mathrm{d}t$, $\mathrm{d}\phi/\mathrm{d}t$, $\mathrm{d}\theta/\mathrm{d}t$), and also the time evolution of the geodesic constants of motion $E$, $L_z$ and $Q$, which are used in  Sec.~\ref{sec:remnantEMRI} to compute the remnant quantities.
The natural frame for EMRIs is defined by the spin of the massive black hole, so both the waveform and trajectories are defined with respect to this frame as illustrated in Fig.~\ref{fig:EMRI}. 
To cohesively append to our existing dataset, it is therefore necessary to transform from this frame to that used by comparable mass simulations,
where the $z$-axis is defined not by the angular momentum of the larger black hole, but by the direction of the orbital angular momentum, or a similar quantity like the orbital angular velocity or Newtonian orbital angular momentum $\vec \omega$, as shown in Fig.~\ref{fig:coorbital}.

\subsection{Creation of a heterogeneous dataset}
\label{sec:blending}
In this work we investigate the properties of the remnant object: the final mass and spin. 
The final mass of the binary is given by
\begin{equation}\label{eq:Mf}
    M_f = 1 - E_{\mathrm{rad}},
\end{equation}
and one can then apply the conservation of angular momentum to compute the final spin $\vec{\chi_f}$:
\begin{align}\label{eq:final_spin}
    M_f^2\; \vec{\chi_f}=m_1^2\; \vec{\chi_1}+ m_2^2\; \vec{\chi_2} + \vec{L}.
\end{align}

For the numerical relativity datasets the final mass and spin were determined from the apparent horizons. In the case of the SXS and BAM dataset they were taken from the supplied metadata, and for our Einstein Toolkit dataset the numbers were averaged over appropriate portions of the late time behaviour of the dataset.
For the extreme mass ratio case, the determination of the final state is described below.

\subsubsection{Choice of reference frame}\label{sec:reference_frame}
%%%%%%%%%%%%%%%%%%%%%%%%%%%%%%%%%%%%%%%%%%%%%%%%%%%%%%%%%%%%%%%%%%%%

In the aligned spin quasi-circular sector, creating a consistent heterogeneous dataset, which combines comparable mass and extreme mass ratio cases, is relatively straightforward. As the spins and orbital plane maintain their orientation as a consequence of equatorial symmetry, 
a natural class of inertial frames can be constructed, where the $z$-axis is the fixed axis of the orbital motion, and the angular coordinate in the orbital plane can be defined based on the separation vector. In the precessing case, this equatorial symmetry of the aligned spin vector is lost; the orbital plane and spin directions are time dependent, and there is in general no natural inertial frame. An intuitive approach to choosing a reference frame is then to work with a non-inertial frame which tracks the precession motion, which drastically simplify the dynamics and waveform  \cite{Tracking,Towards}. This can then be used to construct precessing waveform models in terms of rotating or \emph{``twisting up''} a non-precessing waveform with an inverse rotation that maps a corresponding precessing waveform into the appropriate non-inertial frame \cite{Tracking,Towards}. Unfortunately the natural choices in the EMR limit and comparable mass cases are not equal.

For EMRIs, as discussed above in Sec.~\ref{sec:EMRIs}, fixing an inertial frame by choosing the $z$-axis as the spin axis of the large BH is indeed natural, e.g. it gives rise to a conserved spin component in the $z$-direction of the larger BH, a conserved orbital angular momentum projection $L_z$, and the definition of the conserved inclination $I$ for a fixed geodesic.
In the comparable mass case, this choice has however no meaningful analog, and it has become customary to work in a co-orbital or similar frame, which is defined in the spirit of the quadrupole-aligned frame \cite{Tracking,Towards,minrotcond} for the gravitational wave signal. 
Similar behaviour can be achieved by choosing the $z$-axis as the direction of the orbital angular momentum  $\vec{L}$, or the orbital angular velocity $\vec{\omega}$. 
In the frame defined by the orbital angular momentum, the magnitude of the spin projections parallel and orthogonal to the orbital angular momentum are approximately preserved~\cite{PhysRevD.52.821, TowardsII} and the orbital angular momentum is approximately aligned with the direction of maximum wave emission \cite{Tracking}. In the EMR limit, the projection of the black hole spin onto the orbital angular momentum is preserved, in contrast to the scenario where the orbital angular momentum is replaced by th   e orbital angular velocity $\vec\omega$. 
Some simplification of the gravitational wave signal and dynamics can also be achieved by choosing the $z$-axis as the direction of the total angular momentum  $\vec{J}$, see e.g.~\cite{Towards}, which only varies slowly and by a small amount, except for the case of transitional precession~\cite{Towards}. Furthermore, in the EMR limit, $\vec{J}$ corresponds to the spin of the largest black hole, making it a more suitable choice as the mass ratio increases.

The \NRSur model \cite{NRSur7dq4} uses the quadrupole aligned waveform to define the reference frame, where the $z$-axis is computed as the principal eigenvector of the angular momentum operator as described in~\cite{minrotcond}.
In this work we have chosen to construct our co-orbital frame in terms of orbital quantities, as their formulation is more straightforward than the orbital angular momentum in a numerical relativity evolution.
For some of the ETK waveforms, the gravitational wave signal is too noisy to work with the quadrupole aligned frame, due to an inappropriate configuration of the wave extraction grids. 
This problem and its resolution will be discussed in a separate paper.
The $z$-axis is then chosen to point in the direction of the orbital angular velocity $\vec{\omega}$
\begin{equation}\label{eq:omega}
    \vec{\omega}(t)=\frac{\vec{r} \times \vec{v}}{r^2}=\frac{\vec{r}\times \dot{\vec{r}}}{r^2},
\end{equation}
where $\vec{r}$ is the vector which points from the smallest black hole to the largest ($\vec{r}=\vec{x_1}-\vec{x_2}$), and the $x$-axis is chosen to point in the $\vec{r}$ direction (see Fig.~\ref{fig:coorbital} for visual definition).
The $y$-axis is defined as usual to complete an orthogonal right-handed triad. 

In order to perform this alignment, one needs the time evolution of the two black hole positions, as well as the time evolution of all the quantities we include in our dataset. These are the spins and masses of both initial black holes and the remnant, the emitted waves, the radiated energy and/or orbital frequencies, although additional information on how to get these quantities are recommended to ensure consistency. 
Some public catalogues are hence not adequate for our purposes, e.g. the RIT catalog \cite{RIT4} includes precessing waveforms but it does not provide the trajectory evolution. For future work we also plan to include 
the most recent MAYA catalogue \cite{MAYA2}.

We adopt the same reference time for defining the spin components in a co-orbital frame as in~\cite{NRSur7dq4,NRSur7dq4EmriRemnant}, which is set to $100 M$ before merger. This choice facilitates direct comparisons between our remnant model and \NRSurE \cite{NRSur7dq4EmriRemnant}.
The binary evolution closely approaches the merger state at this reference time, thus one can expect a simpler functional dependence for the remnant quantities.
For the EMR limit, the ISCO provides an approximate plunge time. For our numerical EMRI dataset detailed in Sec.~\ref{sec:EMRIs}, we compare the numerical preserved quantities at merger with those obtained at the ISCO time. For our EMRI dataset, the ISCO time ranges from $-700 M$ for the most anti-aligned cases to $-150 M$ for the aligned ones.
Upon comparing the values for the preserved quantities obtained by solving the geodesic equations at the ISCO with the numerical results, we observe maximum relative errors of 0.1\%. Consequently, we conclude that the ISCO time can be effectively employed as the reference time for the EMR limit without impacting the transition from the comparable mass regime, where the chosen reference time is $-100 M$. 
These observations however suggest that this choice might not be optimal. Instead, a quantity that smoothly transitions from the comparable mass reference time to the ISCO could be more suitable, such as the minimal energy circular orbit (MECO) time \cite{MECO}. 
We leave this as future work, to investigate further a choice of optimal reference time, which allows simple yet precise fits and an accurate match with fast post-Newtonian inspiral codes e.g.,~\cite{spintilts}. These codes serve to bridge the gap between the reference time and some earlier time where waveform models define their spin vectors. 

The rotation of the vector quantities such as the black holes' spins is straightforward, consisting in a fixed rotation to the full time array. 
For the waveforms, it is common to decompose the waves into spherical harmonics and rotate each mode individually via Wigner matrices.
Comparable mass binaries are usually described at the co-orbital frame at some reference time close to the initial time, and performing a fixed rotation to a later co-orbital frame does not suppose higher complications. 
More efforts need to be done however for the extreme mass ratio limit, where the natural frame is defined in terms of the largest black hole spin. From the trajectories one can obtain the vector $\vec{r}=-\vec{x_2}$, as in Eq.~\eqref{eq:x2}, then compute the velocity $\dot{\vec{r}}$ and finally obtain the $z$-axis given by $\hat{\omega}(t_{\mathrm{ref}})$.
In the case of using geodesics, one just inputs the inclination angle at the reference time.
Once the alignment is done for all simulations, we keep the metadata at the chosen reference time -$100\ M$ where the alignment takes place. 
This includes the mass ratio, the two black holes' spins and positions, the reference orbital frequency, the time of merger and finally the remnant quantities.

\subsubsection{Extreme mass ratio limit}\label{sec:remnantEMRI}

In the EMR limit, the final mass and spin can be computed to first order in the mass ratio from the energy and orbital angular momentum at the ISCO, since the contribution of the plunge can be neglected \cite{Kennefick:1995za}. The  quantities $E$ and $L_z$ are preserved for geodesics, and can thus be evaluated directly for any geodesic. This is however not true for the full angular momentum vector, which would require further knowledge about the spacetime.
One can however approximate $L_{\rho}$ by the square root of the Carter's constant $\sqrt{Q}$, and extract the direction of the final spin with respect to the $z$-axis. Since only the $z$ or the in-plane components are preserved, there will be a freedom regarding the in-plane direction of the final spin that we will not be able to fix. Regardless, here we are only interested in the final spin magnitude, so the missing direction does not constrain our work.

Numerically solving the geodesic equations detailed in App.~\ref{app:geodesics} allows one to determine the constants of motion at a specific geodesic. As previously discussed, in the EMR limit, we select the ISCO time as the reference time. At this point, the conditions $R(r_{\mathrm{ISCO}})=R'(r_{\mathrm{ISCO}})=R''(r_{\mathrm{ISCO}})=0$ are satisfied (refer to Eq.~\eqref{eq:R(r)}). Solving this set of algebraic equations provides the constants of motion $(E, L_z, Q)$ at the ISCO, as well as the radius $r_{\mathrm{ISCO}}$.
This numerical procedure is implemented in the \texttt{KerrGeodesics} Mathematica package, which we have utilized extensively to solve precessing geodesics throughout.

For the specific case of non-precessing orbits ($I=0,\pi$), one can derive simple analytical expressions for the energy and angular momentum at the ISCO, given by:
\begin{equation}\label{eq:Eisco_emri}
    \tilde{E}_{\mathrm{ISCO}}(\chi_f) = \sqrt{1-\frac{2}{3\rho_{\mathrm{ISCO}}(\chi_f)}},
\end{equation}
and
\begin{equation}\label{eq:Lisco_emri}
    \tilde{L}_{\mathrm{ISCO}}^{\mathrm{orb}}(\chi_f) = \frac{2 \left( 3\sqrt{\rho_{\mathrm{ISCO}(\chi_f)}}-2\chi_f\right)}{\sqrt{3\rho_{\mathrm{ISCO}}(\chi_f)}},
\end{equation}
where $\rho_{\mathrm{ISCO}}$ is the radius at the ISCO:
\begin{equation*}
    \rho_{\mathrm{ISCO}}(\chi) = 3 + Z_2 - \mathrm{sign}(\chi)\sqrt{(3-Z_1)(3+Z_1+2Z_2)},
\end{equation*}
\begin{equation*}
    Z_1 = 1+(1-\chi^2)^{1/3} \left[(1+\chi)^{1/3}+ (1-\chi)^{1/3}\right],
\end{equation*}
\begin{equation*}
    Z_2 = \sqrt{3\chi^2 + Z_1^2}.
\end{equation*}
From these equations it is clear that the derivative of the final mass and spin with respect to the component spin is singular at $\eta=0$ when the black hole spin is extremal and aligned with the orbital angular momentum. This creates problems when developing a model that covers the entire parameter space, and further work will be required to fully resolve the associated issues.

\begin{figure} 
    \includegraphics[width=1\columnwidth]{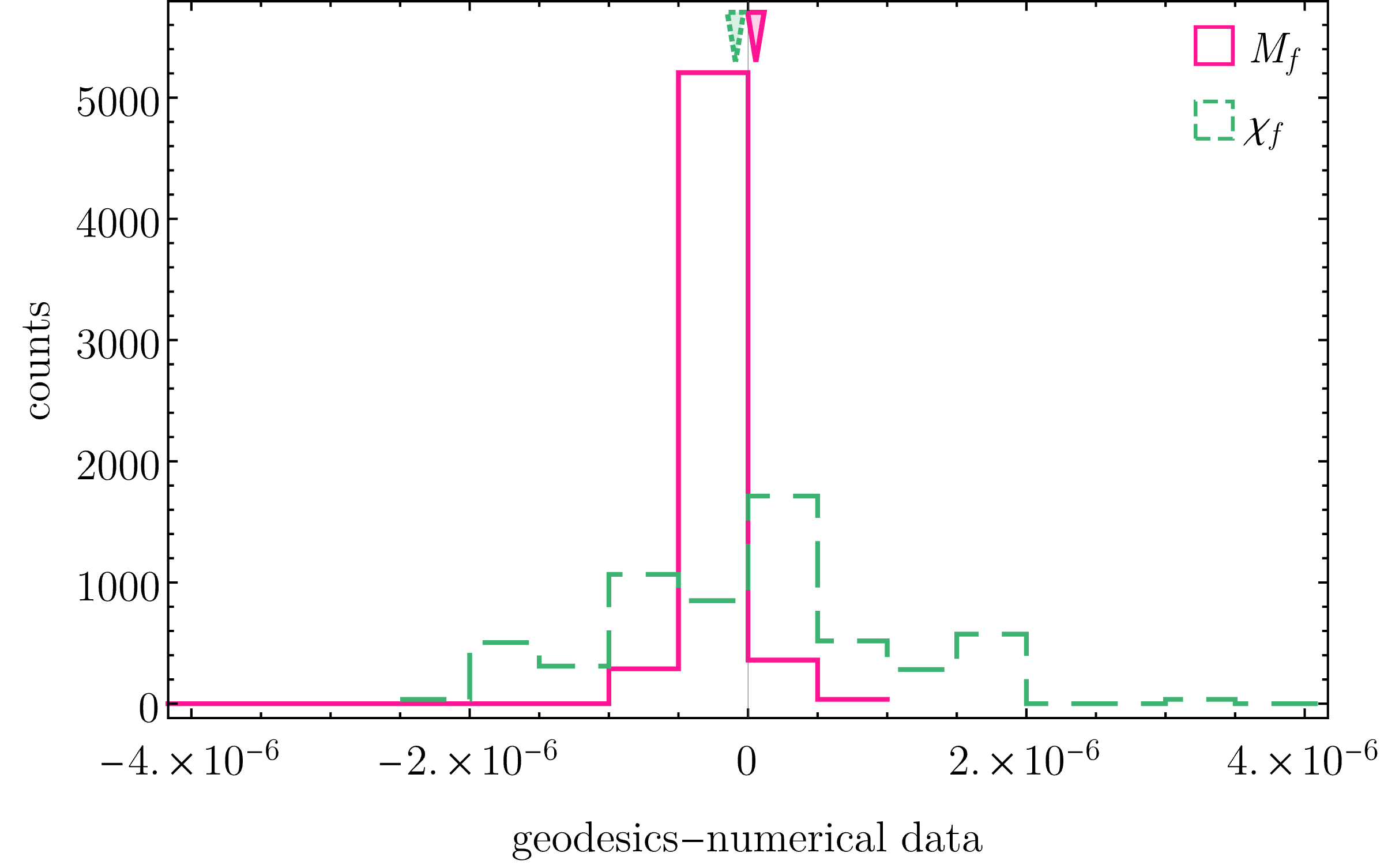}
    \caption{Histogram illustrating the difference between the remnant properties obtained by solving the precessing geodesic equations at the ISCO and the corresponding numerical values from the used EMRI dataset described in Sec.~\ref{sec:EMRIs}. The triangles represent the median value for each distribution.
    \label{fig:comparisongeo}}
\end{figure}

Approximating the orbital angular momentum magnitude $L$ by $\sqrt{L_z^2+Q}$, the remnant quantities in Eqs.~\eqref{eq:Mf} and \eqref{eq:final_spin} depend exclusively on the ``preserved'' quantities $(E,L_z,Q)$. While the numerical dataset provides these values after the plunge, for the geodesic description we take these values from the ISCO and neglect the contribution from the plunge~\cite{Kennefick:1995za}.
We scaled the radiated energy by $\eta$ at linear order, which is consistent with our earlier discussion that the geodesic values are accurate up to order $\eta$.
We compare the final mass and spin magnitude obtained from the precessing geodesic equations and the numerical EMRI data in Fig.~\ref{fig:comparisongeo}, showing a maximum error around $10^{-6}$, comparable to the numerical error expected from the simulations.

\section{Models for the remnant mass and spin}\label{sec:applications}

In this section we develop fits for the remnant mass and spin for precessing binaries, extending some of the ideas which have been used in~\cite{Jimenez-Forteza:2016oae} to create such fits for aligned spin binaries. We use extreme mass ratio data and split the input parameter space (in a hierarchical way) by dimension to design a class of functional forms for the fits. We use information criteria to select the best fit among this class of functions preventing overfitting.

\begin{figure}   
    \includegraphics[width=1\columnwidth]{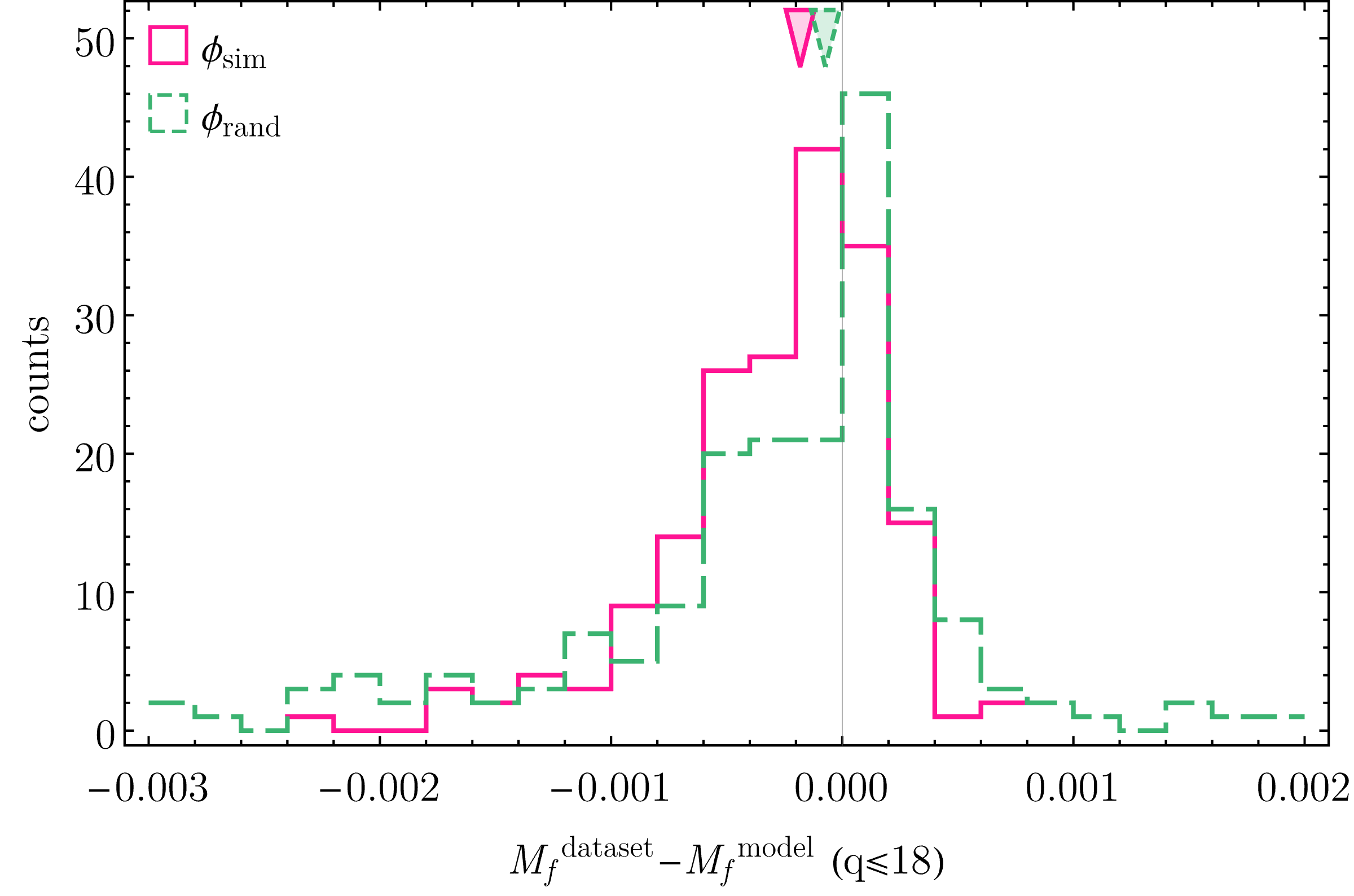}
    \includegraphics[width=1\columnwidth]{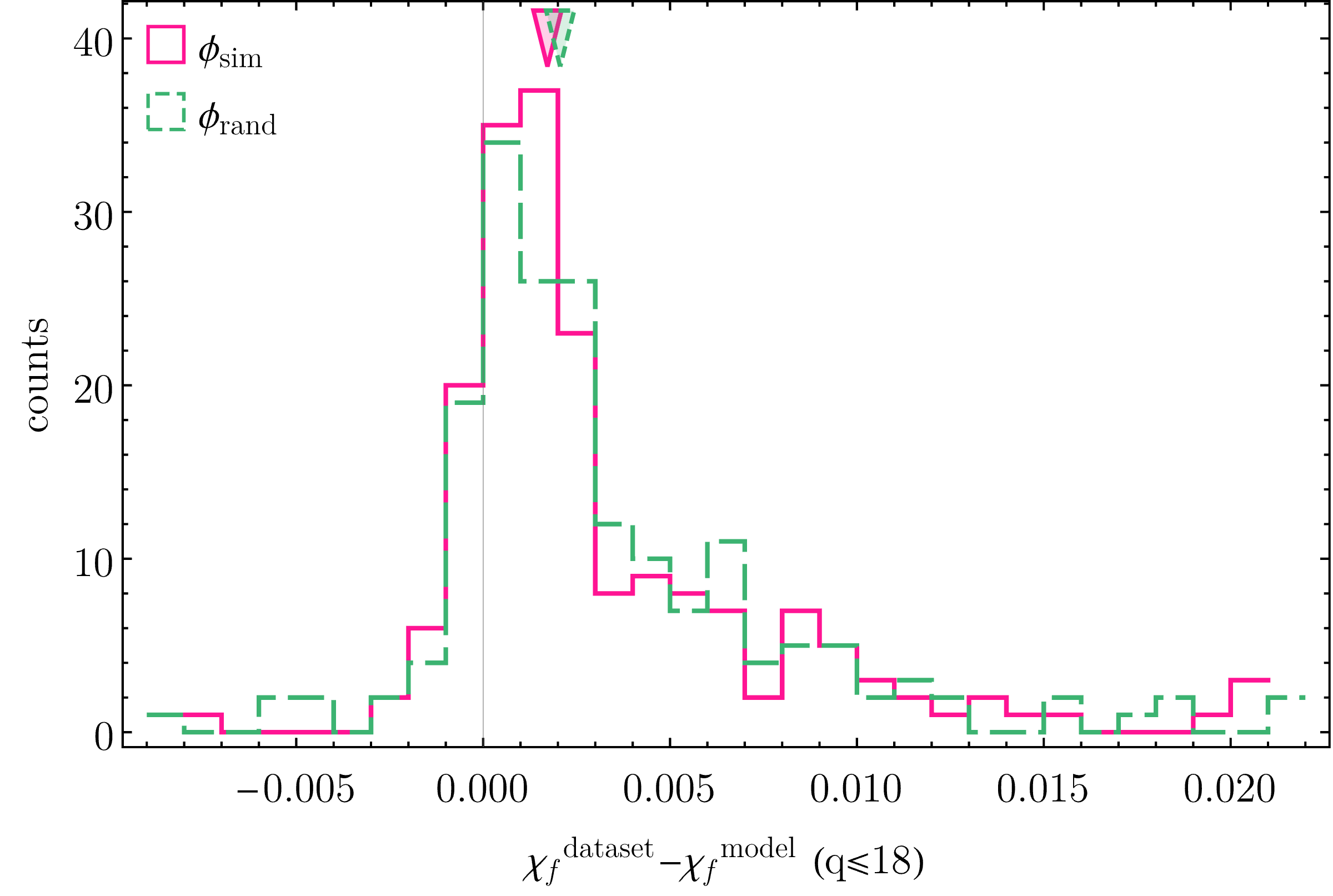}
    \caption{{Error histograms illustrating the effect if the in-plane orientation of the single spin $\phi_1$ on the \NRSurE model for the remnant properties. We consider the corresponding $\phi_1$ of each simulation ($\phi_{\mathrm{sim}}$ in the legend) and a random angle value ($\phi_{\mathrm{rand}}$) and compare the error distributions for our dataset. The top panel displays the errors associated with the final mass (with a RMSE=$5.4\cdot10^{-4}$ for the $\phi_{\mathrm{sim}}$ distribution and RMSE=$8.2\cdot 10^{-4}$ for $\phi_{\mathrm{rand}}$), while the bottom panel depicts the errors related to the final spin magnitude (RMSE($\phi_{\mathrm{sim}}$)=$5.4\cdot10^{-3}$ and RMSE($\phi_{\mathrm{rand}}$)=$5.8\cdot10^{-3}$). Triangles indicate the median value for each distribution.
    Both plots suggest that $\phi_1$ does not significantly influence the error distribution of the model, supporting our decision to exclude that dimension from our studies.}
    \label{fig:phi_dep}}
\end{figure}

At high mass ratios, the contribution of the secondary spin becomes a subdominant effect. This makes the single spin subspace a natural arena when trying to understand the high mass ratio regime for precessing systems and gain intuition on how to bridge the gap to comparable mass binaries (through intermediate mass ratio systems).
The problem thus becomes four-dimensional, with three dimensions due to the largest black hole spin, $(\chi_1,\theta_1,\phi_1)$, and one for the symmetric mass ratio $\eta$.
To assess the impact of the in-plane spin orientation $\phi_1$ on the remnant quantities, we utilize the \NRSurE model, which has been calibrated to numerical relativity and includes the $\phi_1$-dimension. For all configurations in our full numerical relativity single spin dataset we compute the residual error between the NR values for the remnant quantities and those obtained with \NRSurE, first using the value of $\phi_1^{\mathrm{ref}}$ from the simulation, and then a random value. The results, depicted in Fig.~\ref{fig:phi_dep}, reveal that the root mean square errors are virtually unaffected by this additional dimension and that the error distribution exhibits similar behavior.
This supports the decision to exclude the $\phi_1$-dimension in this study, reducing our dimensionality to three free parameters and thereby lowering the computational cost of the procedure. We leave the incorporation of the $\phi_1$ dependence for future work.

The twisting up procedure introduced in Sec.~\ref{sec:blending} permits to understand precession in terms of an approximate map between aligned spin binaries and precessing ones in a co-orbital frame. Our strategy will be to work in a co-orbital frame to facilitate constructing our fits as corrections to the values of the corresponding aligned-spin binary configuration.
The misaligned spin components induce a precessing motion of the binary, which introduces a new timescale compared to aligned-spin systems. During the inspiral this precessing timescale is however much slower than the orbital one, so its effect on the energy radiated in gravitational waves is rather small. For the radiated angular momentum we will see that the situation is slightly more complicated: because the angular momentum and the component spins are time dependent vectors, the final angular momentum is affected by a non-trivial vector addition effect. For a recent discussion in the context of current waveform models see e.g. \cite{Pratten:2020ceb}.

To prevent overfitting we follow Ref.~\cite{Jimenez-Forteza:2016oae}, where aligned spin remnant fits were developed,
and we use the Bayesian Information Criterion (BIC) and Akaike Information Criterion (AIC) as metrics for model selection. These criteria are designed to balance model accuracy and complexity to avoid overfitting.
For further details on the definition of the information criteria see App.~\ref{ap:IC}.
We focus in particular on the BIC, which provides a more restrictive criterion for our purposes. A lower BIC value indicates a more favorable trade-off between model fit and complexity, leading to the selection of a model with improved predictive performance.

Our input dataset consists of the single spin simulations displayed in Fig.~\ref{fig:datasetSS}. Our calibration parameter space extends only up to $\chi_1=0.8$. However, we discuss extrapolation to extreme spins in App.~\ref{app:extrap}, where we conclude that our model extrapolates well to maximally precessing spins. 

We evaluate the accuracy of the remnant fits and compare with the results obtained with \NRSurE, which is calibrated against double spin numerical relativity simulations, and with the remnant fits that are being used in existing precessing phenomenological models \cite{IMRPhenomXPHM, IMRPhenomTPHM}, which are only calibrated to aligned spin simulations.
By subtracting information from the aligned spin sector and EMR before the fits, we can construct simple parameterized  fits which provide a far higher accuracy  that what is currently required for gravitational wave observations, as does the \NRSurE model, but at a much reduced complexity and computational cost. 

\subsection{Remnant mass}\label{sec:finalmass}
\begin{figure}   
    \includegraphics[width=0.8\columnwidth]{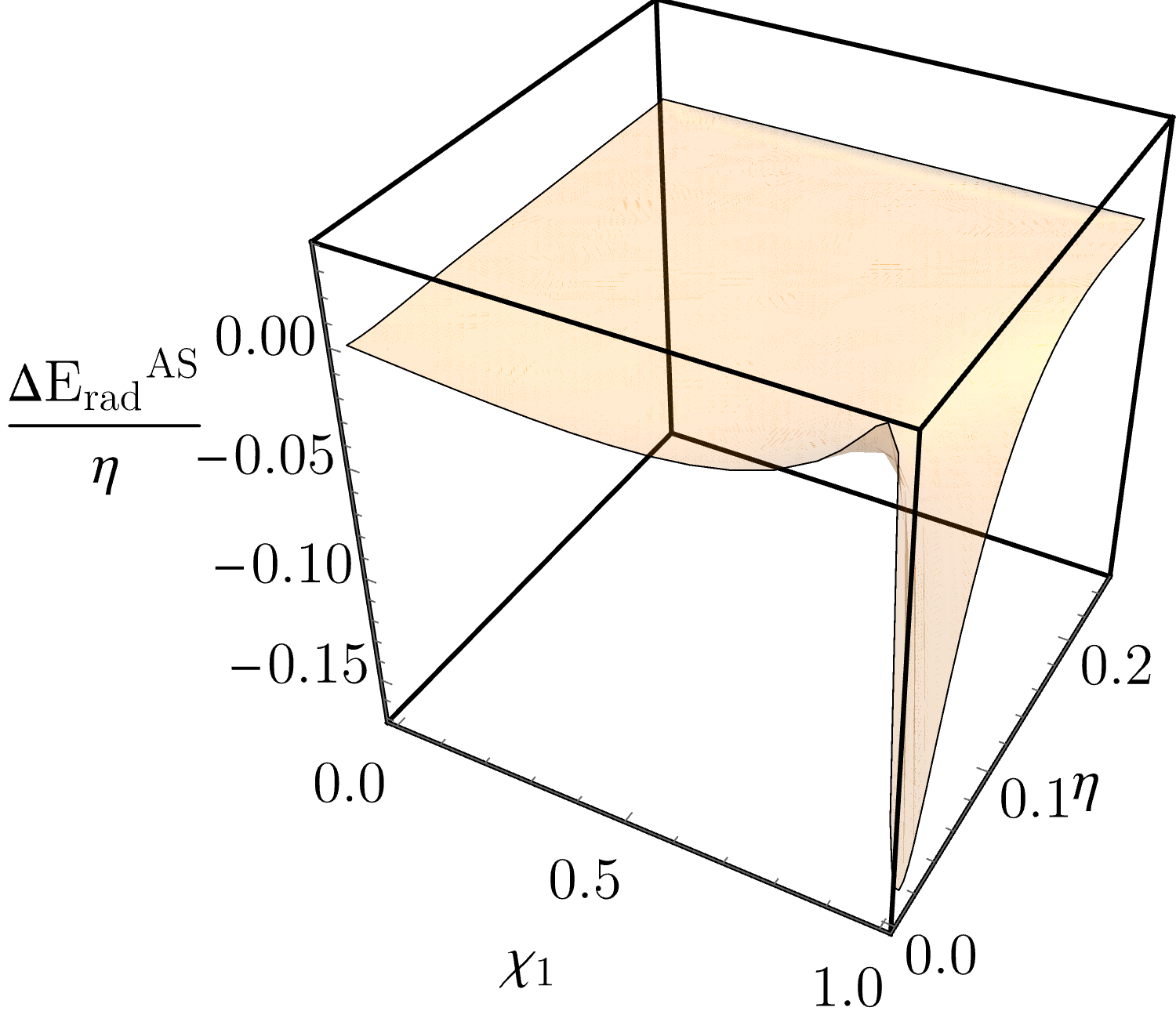}
    \caption{{Difference between the updated aligned spin fit for the radiated energy (see Eq.~\eqref{eq:EradASfit}) and the previous one~\cite{Jimenez-Forteza:2016oae} used in phenomenological families \phX~\cite{phenomxhm} and \phT~\cite{phenomthm} for single spin systems, scaled by the mass ratio. The key enhancement in the new model is a more accurate description of the singularity at $\chi_1 \rightarrow 1$ and $\eta \rightarrow 0$.}
    \label{fig:ERADnew}}
\end{figure}

Since the correction of the final mass due to precession is expected to be small when parameterizing the input spins in a co-orbital frame, it is natural to directly fit the effect of precession on the aligned spin radiated energy.
We then define our fitting quantity as
\begin{equation}
    \Delta E = E^{\mathrm{prec}}_{\mathrm{rad}}\left(\eta,\chi_1,\theta_1\right)-E^{\mathrm{AS}}_{\mathrm{rad}}\left(\eta,\chi_1\cos(\theta_1),\chi_2\cos(\theta_2)=0\right),
    \label{eq:deltaE}
\end{equation}
so the final mass of a precessing binary will be modified as
\begin{equation}
    M_f=1-E_{\mathrm{rad}}=1-\left(E_{\mathrm{rad}}^{\mathrm{AS}}(\eta,\chi_1 \cos(\theta_1))+\Delta E(\eta,\chi_1,\theta_1)\right),
    \label{eq:MffromErad}
\end{equation}
where all the input parameters are taken at the reference time.
Since by design our fitted quantity $\Delta E$ has very small values, we need to ensure that poor accuracy of the aligned spin fit for the energy does not contaminate our error budget. 
What we found is that the previous fit developed in Ref.~\cite{Jimenez-Forteza:2016oae} did not accurately capture the radiated energy close to the joint extremal spins and extreme mass ratio limit. Hence, we first improved the aligned spin fit near this singular point as shown in Fig.~\ref{fig:ERADnew}, where we defined $\Delta E_{\mathrm{rad}}^{\mathrm{AS}}$ as the difference between the updated fit and the old one. The updated expression is provided in App.~\ref{ap:fits}, Eq.~\eqref{eq:EradASfit}, and will be discussed in detail elsewhere.

\begin{figure}   
    \hspace*{-2.5cm}\includegraphics[width=0.73\columnwidth]{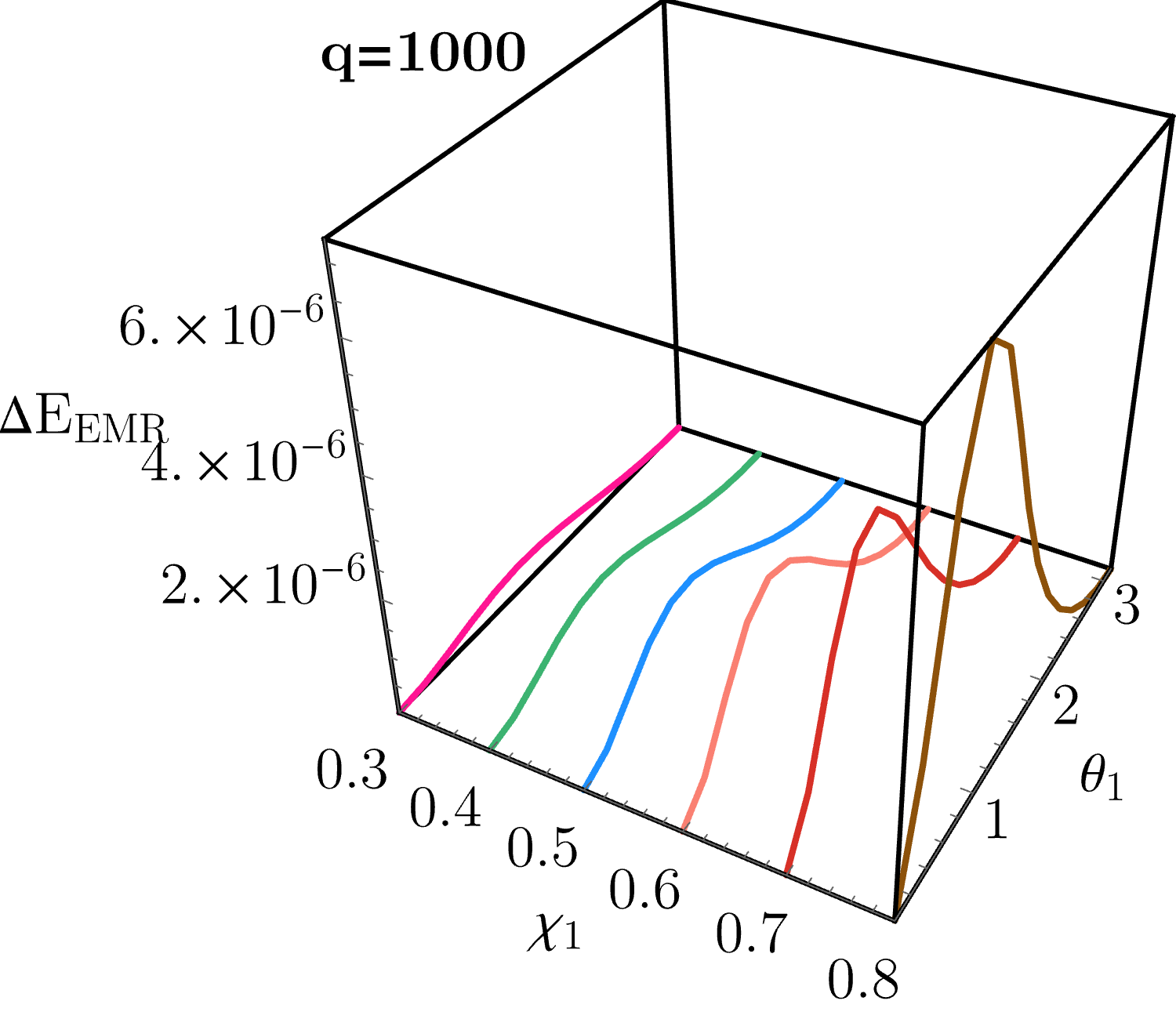}
    \includegraphics[width=0.97\columnwidth]{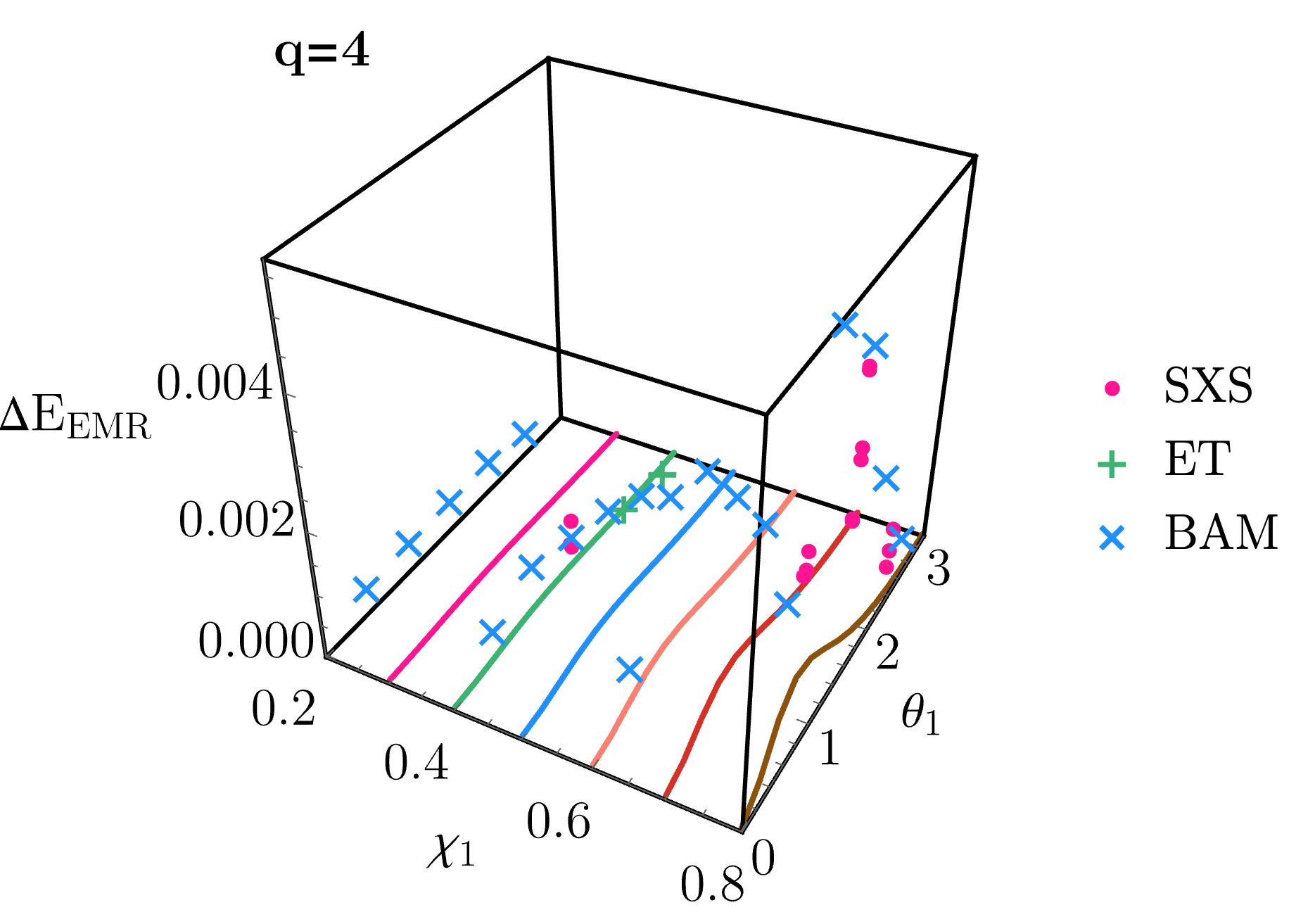}
    \caption{{Numerical evaluations of $\Delta E$ as defined in Eq.~\eqref{eq:deltaE} for the extreme mass ratio limit (denoted as $\Delta E_{\mathrm{EMR}}$) at a fixed mass ratio while varying the black hole's spin magnitude $\chi_1$ and its orientation with respect to the orbital frequency at the reference time $\theta_1$. 
    The precessing and aligned spin radiated energies are obtained from the geodesic equations of motion, which provide the linear contribution in $\eta$ to the energy. The precessing radiated energy at the ISCO is obtained with the \texttt{KerrGeodesics} package, whereas the aligned spin energy is computed from Eq.~\eqref{eq:Eisco_emri}. 
    The top panel corresponds to a mass ratio of 1000, where the geodesic equations are expected to be valid, while the lower panel corresponds to a mass ratio of 4. In the lower plot, we included the single spin simulations from Fig.~\ref{fig:datasetSS} that fall into this subspace.
    \label{fig:deltaEEMRI}}}
\end{figure}

In order to improve the conditioning of our fitting method for the EMRI regime we use the geodesic results, which are valid at linear order in $\eta$:
We subtract the resulting linear-in-$\eta$ term from the numerical dataset and only include higher powers of $\eta$ in our fits.
Fig.~\ref{fig:deltaEEMRI} shows that the linear-in-$\eta$ term is small for comparable masses, which benefits this strategy by not contributing much where the linear approximation is not valid.
We therefore fit the quantity  $\overline{\Delta E}$ defined as
\begin{equation}
    \overline{\Delta E} = \Delta E - \Delta E_{\mathrm{EMR}}.
    \label{eq:deltaEfinal}
\end{equation}
Here $\Delta E_{\mathrm{EMR}}$ is computed using the \texttt{KerrGeodesics} Mathematica package for the energy in the precessing case and Eq.~\eqref{eq:Eisco_emri} for the aligned spin energy case. 

\begin{figure}   
    \includegraphics[width=1\columnwidth]{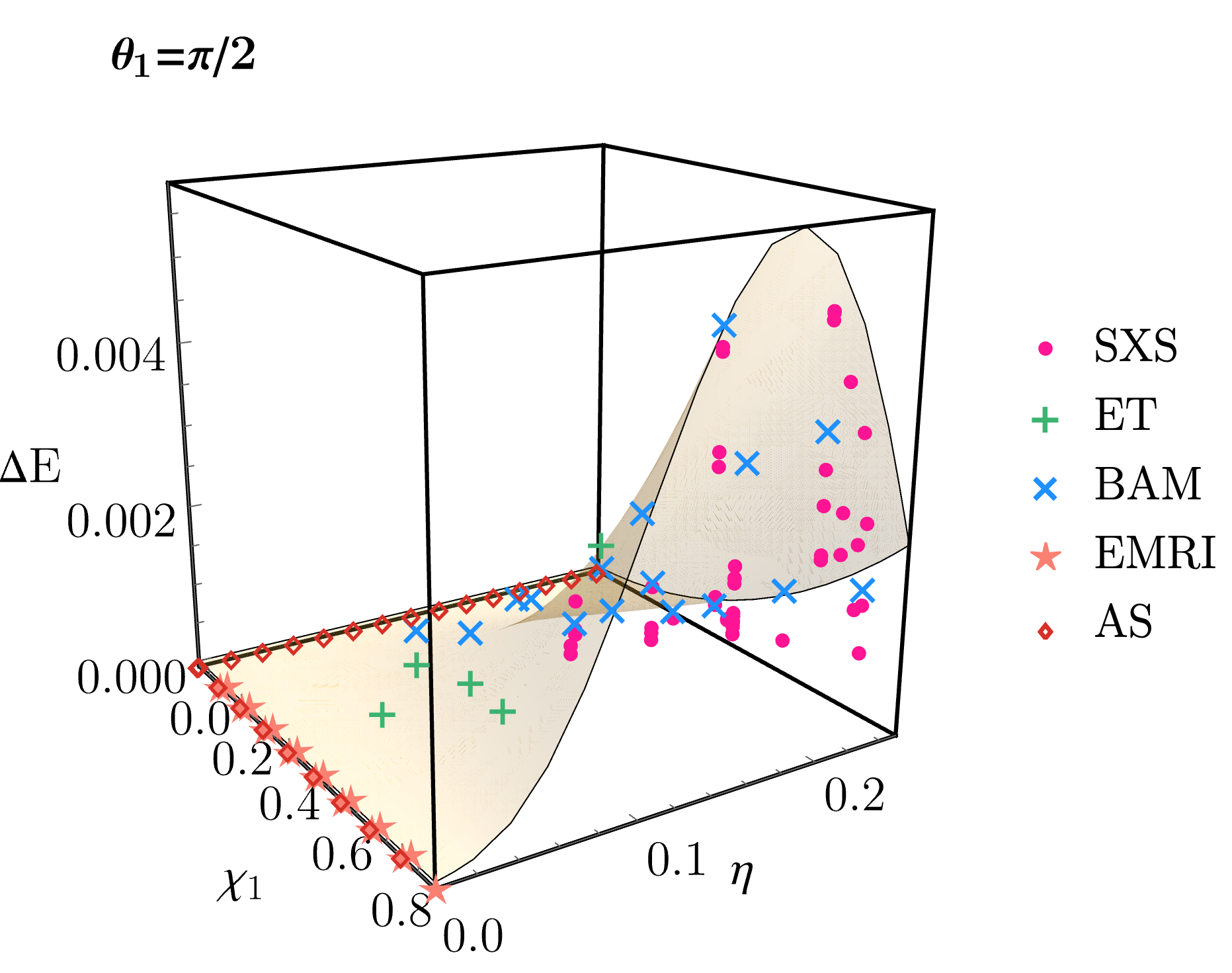}
    \caption{{Numerical evaluation of $\Delta E$ as defined in Eq.~\eqref{eq:deltaE}, obtained from the parameterized fit $\overline{\Delta E}$~\eqref{eq:deltaEfit} and $\Delta E_{\mathrm{EMR}}$, at a fixed spin orientation $\theta_1=\pi/2$, while varying the mass ratio $\eta$ and the spin magnitude $\chi_1$. The figure includes the single spin precessing simulations shown in Fig.~\ref{fig:datasetSS} that fall into this subspace.}
    \label{fig:E_full_data}}
\end{figure}

In order to develop a suitable ansatz for a parametric fit across the three-dimensional space $(\eta, \chi_1,\theta_1)$
we first visualize  only two dimensions and show results for fixed values of $\theta_1$, chosen as the evenly spaced grid points of the BAM catalogue \cite{Cardiff-catalog}: $\theta_1 \in \left\{\frac{\pi }{6},\frac{\pi }{3},\frac{\pi }{2},\frac{2 \pi }{3},\frac{5 \pi }{6}\right\}$.
This way we ensure that each fitted surface contains numerical relativity data.
By definition, $\overline{\Delta E}$ (as well as $\Delta E$) has to vanish at the boundaries $\theta_1 \in \{0,\pi\}$, so that we recover the aligned and anti-aligned limits.
The simple structure of the numerical values of $\Delta E$ (and hence $\overline{\Delta E}$) across the $\eta-\chi_1$ subspace at the fixed values of $\theta_1$ (see e.g. Fig.~\ref{fig:E_full_data}) suggests that a simple polynomial ansatz can effectively capture its behaviour.
More specifically, our ansatz consists of $\eta^a\chi_1^b$-like terms using a rectangular grid in $(a,b)$. Visually inspecting the data, the highest order fit that avoids overfitting for any fixed $\theta_1$ is given by $a\leq 5$ and $b \leq 2$, which results in 18 terms in the polynomial ansatz. 
However, many of these terms can be discarded.

\begin{figure}   
    \includegraphics[width=1\columnwidth]{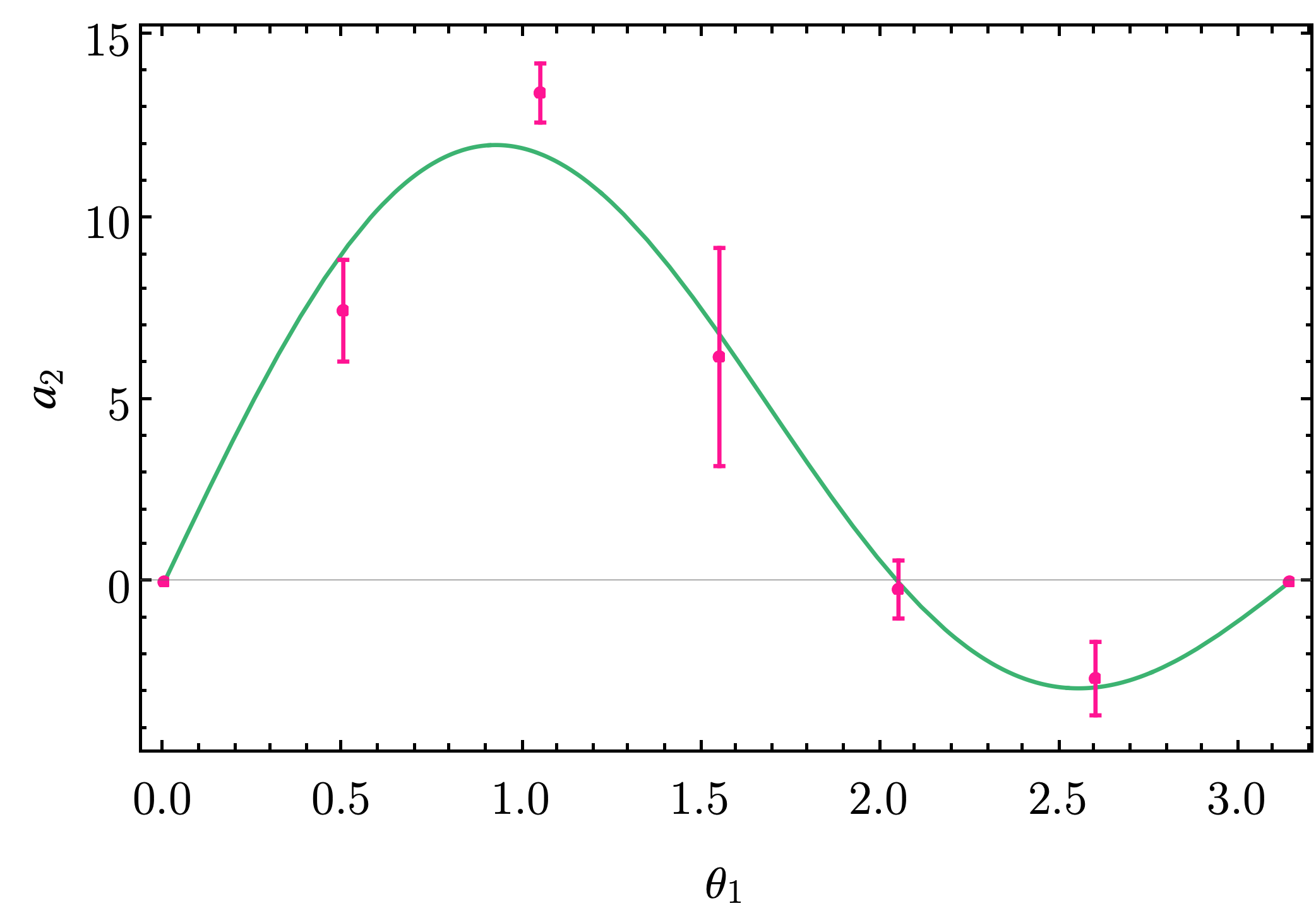}
    \caption{{Functional dependence of the $\eta^3 \chi_1$ term on the angle $\theta_1$ (following the ansatz given by Eq.~\eqref{eq:aiE}), as depicted in Eq.~\eqref{eq:deltaEfit}. Similar results are obtained for the remaining 5 terms of the parameterized fit for $\overline{\Delta E}$. All fits were performed with the \texttt{Fit} function in Mathematica \cite{fit}.}
    \label{fig:E_coeffs}}
\end{figure}

We set the constant term ($a=b=0$) of the expansion to $0$ because, at $\chi_1=0$ and $\eta\rightarrow 0$, it holds that $\overline{\Delta E}=0$, leaving us with 17 terms.
We utilize the \texttt{LinearModelFit} function from Mathematica~\cite{linearmodelfit} to fit the numerical data for each $\theta_1$ surface and record the BIC of the resulting model across all $\theta_1$s. 
Subsequently, we perform a weighted averaging of the BIC for each surface, assigning weights of $0.05$ for $\theta_1 \in \{\frac{\pi}{6}, \frac{5\pi}{6}\}$, of $0.2$ for $\theta_1\in \{\frac{\pi}{3}, \frac{2\pi}{3}\}$, and finally, $0.5$ for $\theta_1=\frac{\pi}{2}$.
The assignment of weights depends on the nature of the quantity being fitted: the magnitude is more significant for highly precessing systems, making the results more reliable in the region closer to in-plane spins ($\theta_1 \sim \pi/2$). Conversely, for systems close to aligned or anti-aligned configurations, the value is so small that it is overshadowed by the numerical error of the simulations. By employing weighted averaging, we ensure that the fitting procedure is not dominated by the numerical errors in our dataset, while still taking into account all cases.
This process is repeated iteratively for a modified ansatz where each term is removed, saving the averaged BIC for each case. 
We retain the ansatz with the lowest mean BIC only if it falls below the BIC value of the initial polynomial set. This procedure is reiterated until removing more terms no longer contributes positively to the final fit. With this procedure, we obtain a final fit with only 7 terms favoured by about -44 in relative BIC. Then, for every value of $\theta_1 \in \left\{\frac{\pi }{6},\frac{\pi }{3},\frac{\pi }{2},\frac{2 \pi }{3},\frac{5 \pi }{6}\right\}$, we have
\begin{equation}
     \{a_i\}_{i=1}^{i=7}\left(\eta ^2 \chi_1 ,\eta ^3 \chi_1 ,\eta ^3 \chi_1^2,\eta ^4 \chi_1 ,\eta ^4 \chi_1 ^2,\eta ^5 \chi_1 ,\eta ^5 \chi_1 ^2\right).
    \label{eq:fitE1}
\end{equation}
Upon visually inspecting each $a_i$, we observed that the contribution of the $\eta^2 \chi_1$ term was minimal, supporting its removal from the final fit.
Consequently, we are left with 6 coefficients to fit for $\theta_1$. 
Thereafter, we proceed to fitting the $\theta_1-$dependence of the coefficients $a_i$.
As previously mentioned, at the boundaries $\theta_1 \in \{0,\pi\}$, we recover the aligned and anti-aligned limits, implying that $a_i(0)=a_i(\pi)=0$. Given that $\theta_1$ is an angle, it makes sense to propose a sinusoidal ansatz such as
\begin{equation}\label{eq:aiE}
a_i(\theta_1)=A_i\sin{\theta_1}+B_i\sin{2\theta_1}, 
\end{equation}
which will always satisfy the boundary conditions.
Figure~\ref{fig:E_coeffs} shows the functional dependence of the $\eta^3\chi_1$ term in $\theta_1$ and the corresponding fit from the above ansatz in Eq.~\eqref{eq:aiE}.

The final expression for $\overline{\Delta E}(\eta,\chi_1,\theta_1)$ is given by
\begin{equation}
\begin{split}
    \overline{\Delta E}(\eta,\chi_1,\theta_1) =\ 
    &\eta ^3 \chi_1 \left[0.759123 \sin (\theta_1 )-2.33392 \sin (2 \theta_1 )\right]+\\
    &\eta ^3 \chi_1^2 \left[6.51059 \sin (\theta_1 )+7.06906 \sin (2 \theta_1 )\right]+\\
    &\eta ^4 \chi_1 \left[-11.7873 \sin (\theta_1 )+22.364 \sin (2 \theta_1 ) \right] +\\
    &\eta ^4 \chi_1^2 \left[-37.0594 \sin (\theta_1 )-63.3841 \sin (2 \theta_1 ) \right] +\\
    & \eta ^5 \chi_1 \left[35.0427 \sin (\theta_1 )-51.36 \sin (2 \theta_1 )\right] +\\
    & \eta^4 \chi_1^2 \left[-37.0594 \sin (\theta_1 )-63.3841 \sin (2 \theta_1 )\right].
    \label{eq:deltaEfit}
\end{split}
\end{equation}
Figure~\ref{fig:E_full_data} shows $\Delta E$ computed as in Eq.~\eqref{eq:deltaEfinal} together with the numerical values for the entire dataset at $\theta_1=\pi/2$.
This fit can now be inserted into Eq.~\eqref{eq:MffromErad} in order to get the new model for the mass of the remnant object. To compute the final mass of the corresponding aligned system, we rely on the updated \phX model given by Eq.~\eqref{eq:EradASfit}.

We can now assess the accuracy of our new model. For the entire single spin precessing dataset, we calculate the final mass using our model (Eqs.~\eqref{eq:MffromErad},~\eqref{eq:deltaEfit}), denoted as \texttt{PhenNew}. We compare this with the current \phX model, which does not account for the $\Delta E$ correction (\texttt{PhenXP}), and with \texttt{NRSur7dq4EmriRemnant} for validation. 
\begin{figure}   
    \includegraphics[width=1\columnwidth]{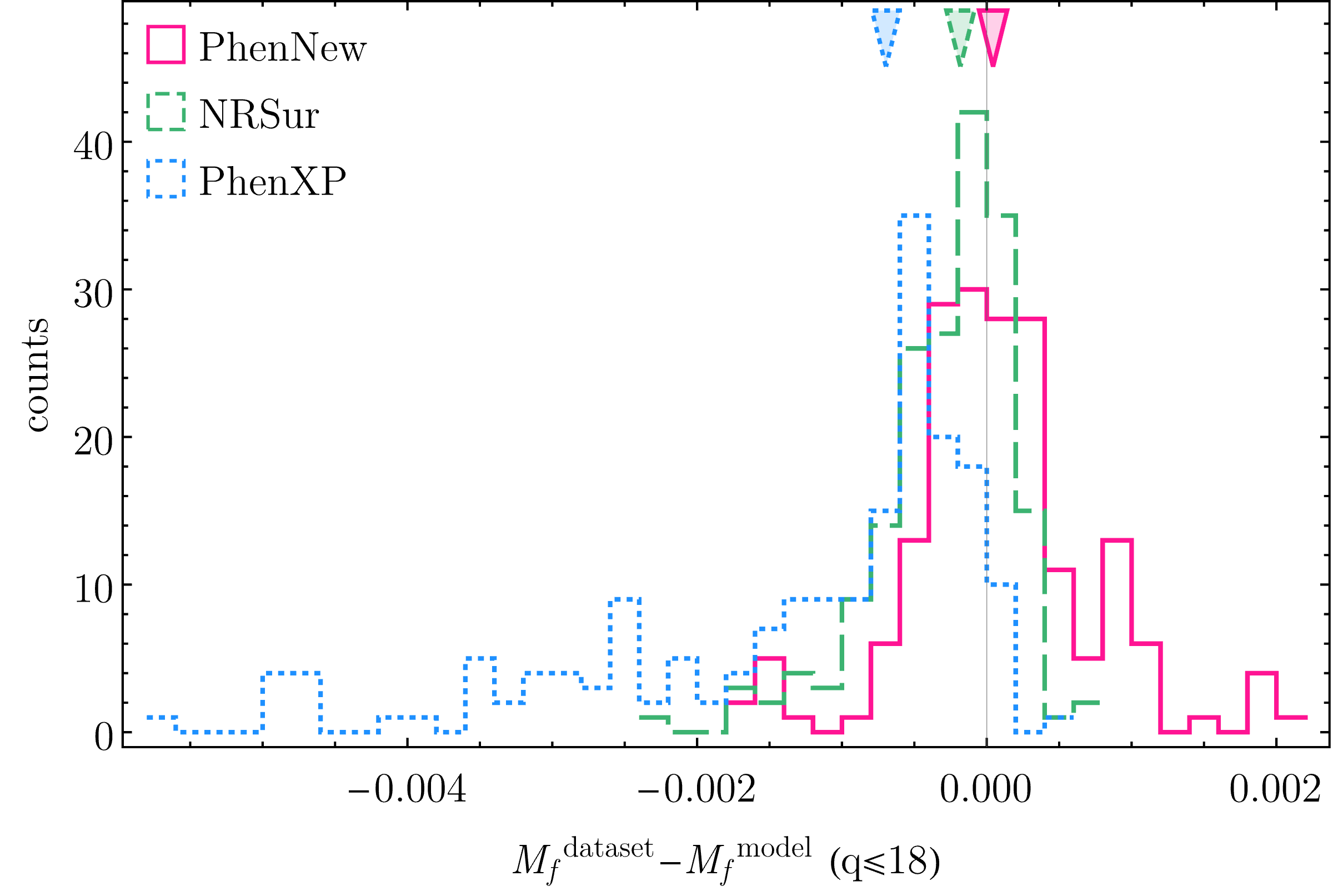}
    \caption{{Histogram of the errors in the remnant mass computed with each of the three models for our single spin precessing dataset presented in Fig.~\ref{fig:datasetSS}. 
    The model developed in this project is labeled as \texttt{PhenNew}, the underlying model as \texttt{PhenXP} and \NRSurE as \texttt{NRSur}. 
    The triangles above the distributions represent their median values, also included in Table~\ref{table:errorsmodel}. This table provides additional information on the distributions: the computational time needed to evaluate the dataset for each model, along with the root mean square errors (RMSEs). Note that the \texttt{PhenXP} model shows a sensible bias that is significantly reduced in the \texttt{PhenNew} model.}
    \label{fig:errorsE}}
\end{figure}
Figure~\ref{fig:errorsE} presents the histogram of errors associated with each model. Results are based on the NR data from our dataset, comprising 184 simulations with $q \leq 18$. The EMRI dataset has been excluded from the comparison due to its large number of simulations and small errors.
Table \ref{table:errorsmodel} provides the computational time required to evaluate the final masses and spins for the whole NR dataset, along with the numerical values of the median error and root mean square error (RMSE) computed as in Eq.~\eqref{eq:RMSE}.
The \texttt{PhenXP} model for the final mass $M_f$ involves evaluating the aligned-spin model for the radiated energy using Eq.~\eqref{eq:EradASfit}. On the other hand, \texttt{PhenNew} evaluates both Eq.~\eqref{eq:EradASfit} and the parameterized fit for $\overline{\Delta E}$~\eqref{eq:deltaEfit}, as well as $\Delta E_{\mathrm{EMR}}$, using the \texttt{KerrGeodesics} package. Note that computational times refer to a straightforward implementation in Mathematica, with most of the computational time required to solve the precessing geodesic equations to obtain $\Delta E_{\mathrm{EMR}}$. An optimized implementation, and a fit to the analytically known EMR results would dramatically accelerate the evaluation.
For the evaluation of the \texttt{NRSur} model, we utilized the \textsc{SurfinBH} python package \cite{NRSur7dq4} with the \NRSurE \cite{NRSur7dq4EmriRemnant} model. In this case, the evaluation time is provided as a single number for both the final mass and spin because both are returned together as an array.
\begin{table}
\begin{tabular}{ccccc}
\cline{3-5}
\noalign{\vskip\doublerulesep
         \vskip-\arrayrulewidth}
\cline{3-5}       
 & & \texttt{PhenNew} & \texttt{PhenXP} & \texttt{NRSur} \\ \hline \hline
\multirow{2}{*}{$M_f$} & Median & \, $4.4\cdot 10^{-5}$ \, & \, $-7.0\cdot 10^{-4}$ \, & \, $-1.8\cdot 10^{-4}$ \, \\
& RMSE & $6.4\cdot 10^{-4}$ & $1.8\cdot 10^{-3}$ & $5.4\cdot 10^{-4}$ \\ \hline
\multirow{2}{*}{$\chi_f$} & Median & $1.5\cdot 10^{-4}$ & $-8.6\cdot 10^{-3}$ & $1.7\cdot 10^{-3}$ \\
& RMSE & $3.4\cdot 10^{-3}$ & $1.5\cdot 10^{-2}$ & $5.4\cdot 10^{-3}$ \\ \hline
\multicolumn{2}{c}{Evaluation time (s)} & 0.10/0.5 & 0.005/0.01 & 112 \\ \hline \hline
\end{tabular}    
\caption{Median value and root mean square error (RMSE) of the error distributions of the remnant mass $M_f$ and spin magnitude $\chi_f$ for different models, with respect to the numerical relativity dataset shown in Fig.~\ref{fig:datasetSS}. The histograms of the distributions are shown in Figs.~\ref{fig:errorsE} and \ref{fig:errorschi}, respectively.
The last row provides the time needed to evaluate the complete dataset with each model for the final mass/final spin. For the \NRSurE model (\texttt{NRSur}) only one value is provided because their python implementation jointly returns both quantities.
\label{table:errorsmodel}}
\end{table}

These findings demonstrate that our new model for the final mass achieves an accuracy comparable to the \NRSur model, surpassing the original \texttt{PhenXP} model, while maintaining its computational efficiency. Additionally, \texttt{PhenNew} exhibits a less biased error distribution compared to the old model, which tends to overestimate the final mass.

To ensure completeness, we assessed the resulting model beyond our calibration region, specifically when $\chi_1 > 0.8$. We focused on scenarios where precession effects are maximal, corresponding to an in-plane spin configuration, i.e. $\theta_1=\pi/2$. 
We include the result of the extrapolation in App.~\ref{app:extrap}.
Despite the absence of numerical data in that region, the study indicates that the extrapolation behaves well, and no dubious features emerge outside the calibration regime.

\subsection{Remnant spin}\label{sec:finalspin}
%%%%%%%%%%%%%%%%%%%%%%%%%%%%%%%%%%%%%%%%%%%%%%

If one assumes the twisting-up approximation and that the in-plane and aligned spin components are conserved, then one can write the final spin magnitude as
\begin{equation}
\chi_f^{\mathrm{prec}}=\sqrt{{\chi_f^{\mathrm{AS}}}^2+\frac{m_1^4}{M_f^4}{\chi_\perp^2}},
    \label{eq:deltasqQA}
\end{equation}
where $\chi_\perp$ is the total in-plane spin.
Variants of this approximation with different assumptions to compute $\chi_\perp$ have been used in the \phX and \phT waveform models \cite{IMRPhenomXPHM,IMRPhenomTPHM}.
In our case, considering only the single spin sector, $\chi_\perp$ simply becomes the in-plane component of the larger black hole.
However, the in-plane and orthogonal spin components are not exactly conserved, and we therefore introduce a correction term $\delta^2$ that we fit to our numerical dataset. Note that denoting the unknown quantity as $\delta^2$ is an abuse of notation, since it is not necessarily positive, and indeed typically is negative. 
Consequently, the final spin of a precessing system can be written as
\begin{equation}
\chi_f^{\mathrm{prec}}=\sqrt{{\chi_f^{\mathrm{AS}}}^2+\frac{m_1^4}{M_f^4}{\chi_\perp^2+\delta^2}},
    \label{eq:deltasq}
\end{equation}
where $\chi_1^{\perp}$ in the single spin case is given by $\chi_1^{\perp}=\chi_1\sin(\theta_1)$, and $M_f= 1-E_{\mathrm{rad}}$, where we use the model of the previous section to compute $E_{\mathrm{rad}}$.

\begin{figure}   
    \hspace*{-2cm}\includegraphics[width=0.65\columnwidth]{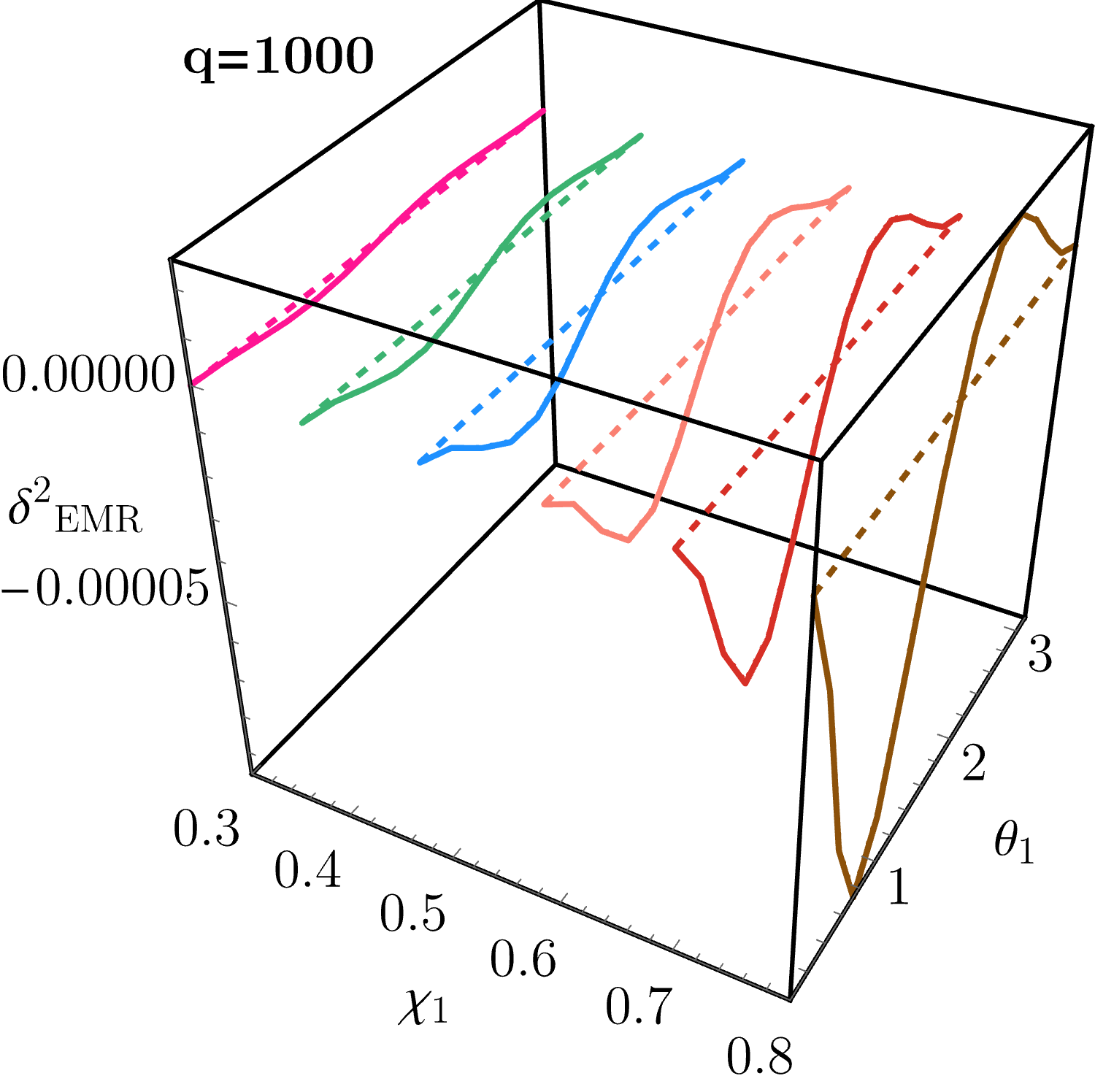}
    \includegraphics[width=0.97\columnwidth]{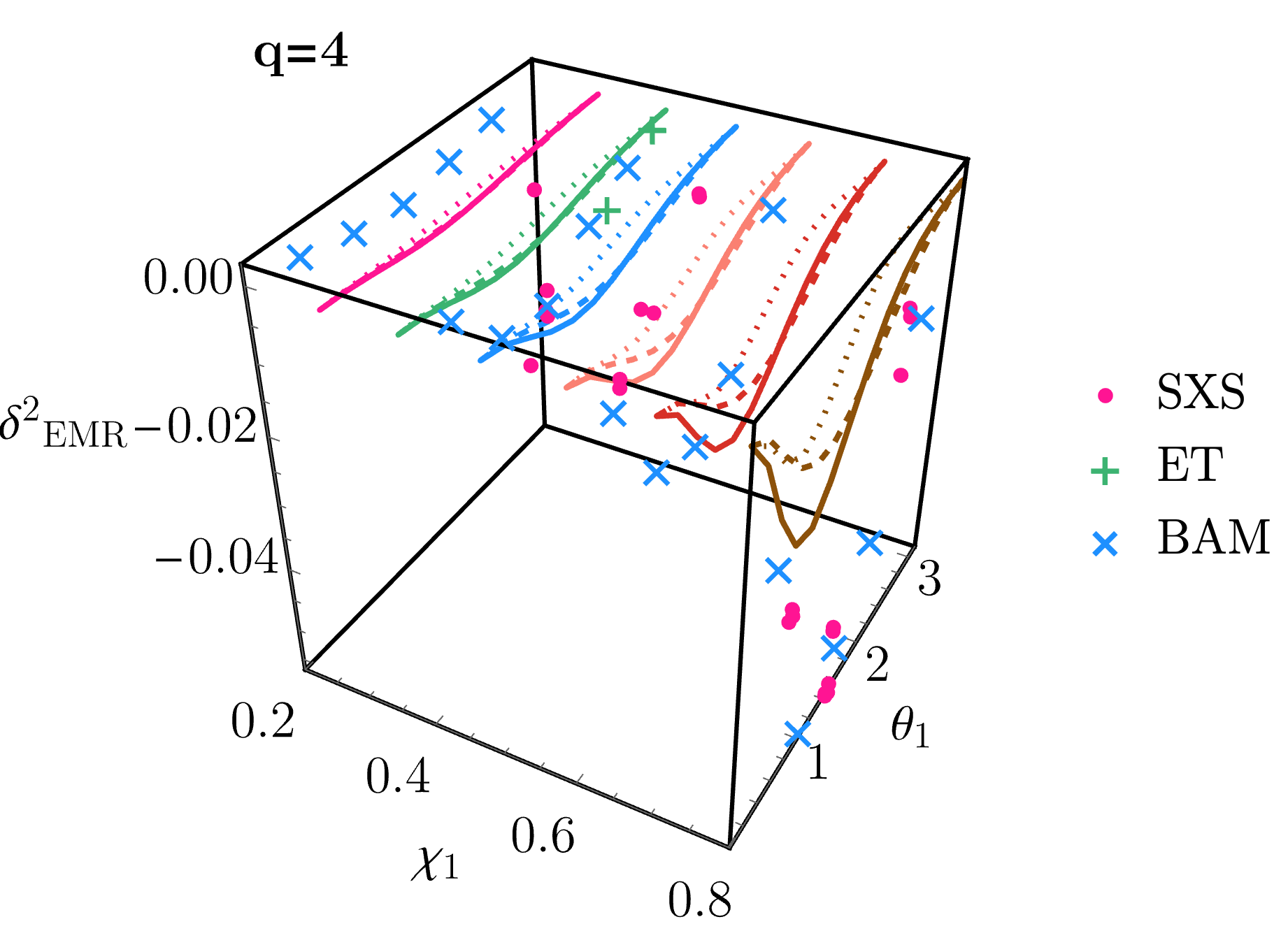}
    \caption{{Numerical evaluation of $\delta^2_{\mathrm{EMR}}$ as defined in Eq.~\eqref{eq:deltasqEMRI}. The constants of motion $L_z$, $Q$ and $E$ are obtained from the precessing geodesic equations, while $L_z^{\parallel}$ is computed from Eq.~\eqref{eq:Lisco_emri}. Dotted lines show the linear term in Eq.~\eqref{eq:deltasqEMRI}} and dashed lines, the quadratic term.
    We solve the precessing geodesic equations at the ISCO using the \texttt{KerrGeodesics} package at a fixed mass ratio, while varying the black hole's spin magnitude $\chi_1$ and its orientation with respect to the orbital frequency at the reference time $\theta_1$. 
    The top panel corresponds to a mass ratio of 1000, where the geodesic equations are expected to be valid, while the lower panel corresponds to a mass ratio of 4, where they are expected to fail. In the lower plot, we included the single spin simulations from Fig.~\ref{fig:datasetSS} that fall into this subspace. 
    \label{fig:deltaEMRI}}
\end{figure}

One can develop the previous equation and turn it into a closed-form approximation for the extreme mass ratio limit by using Eq.~\eqref{eq:final_spin} for both the precessing and aligned final spin, assuming that the Carter's constant is approximately the in-plane orbital angular momentum ($Q \approx L_{\rho}^2$). This procedure results in
\begin{equation}
\begin{split}
    \delta^2_{\mathrm{EMR}}=\frac{M^4}{M_f^4}&\left[\frac{m_2^2}{m_1^2}\left(\tilde{L_z}^2+\tilde{Q}-\tilde{L_z^{\parallel}}\right)+ \right. \\
    & \left. 2\frac{m_2}{m_1}\chi_1\cos(\theta_1)\left(\sqrt{\tilde{L_z}^2+\tilde{Q}}-\tilde{L_z^{\parallel}}\right) \right].
    \label{eq:deltasqEMRI}
\end{split}
\end{equation}
The main advantage of this expression lies in the fact that it relies exclusively on geodesic information for precessing and aligned equations. Remarkably, even for close to comparable masses, its behavior closely resembles that obtained from numerical data, up to a scaling factor, as can be seen in the lower panel of Fig.~\ref{fig:deltaEMRI}. 
Equation~\eqref{eq:deltasqEMRI} consists of two contributions: the first term, quadratic in $1/q$, and the second term, linear. 
Both terms are shown in Fig.~\ref{fig:deltaEMRI}. 
The linear term dominates for extreme mass ratios (in dotted lines, covered by the continuous lines) and exhibits oscillations due to the cosine dependence of the inclination angle. However, as the mass ratio increases, these oscillations are overshadowed by the growth of the quadratic contribution (depicted by dashed lines), as shown in the lower panel of Fig.~\ref{fig:deltaEMRI}. It turns out that in order to reproduce our numerical data in the comparable mass regime it is best to keep both the linear and quadratic in $\eta$ terms.
Following the same motivation as for the energy, we proceed to subtract $\delta^2_{\mathrm{EMR}}$ from our fitting quantity $\delta^2$ to capture the EMRI regime, defining
\begin{equation}
    \overline{\delta^2}=\delta^2-\delta^2_{\mathrm{EMR}},
    \label{eq:deltaEMRIfin}
\end{equation}
where again $\delta^2_{\mathrm{EMR}}$ is computed using the \texttt{KerrGeodesics} Mathematica package.

\begin{figure}   
    \includegraphics[width=1\columnwidth]{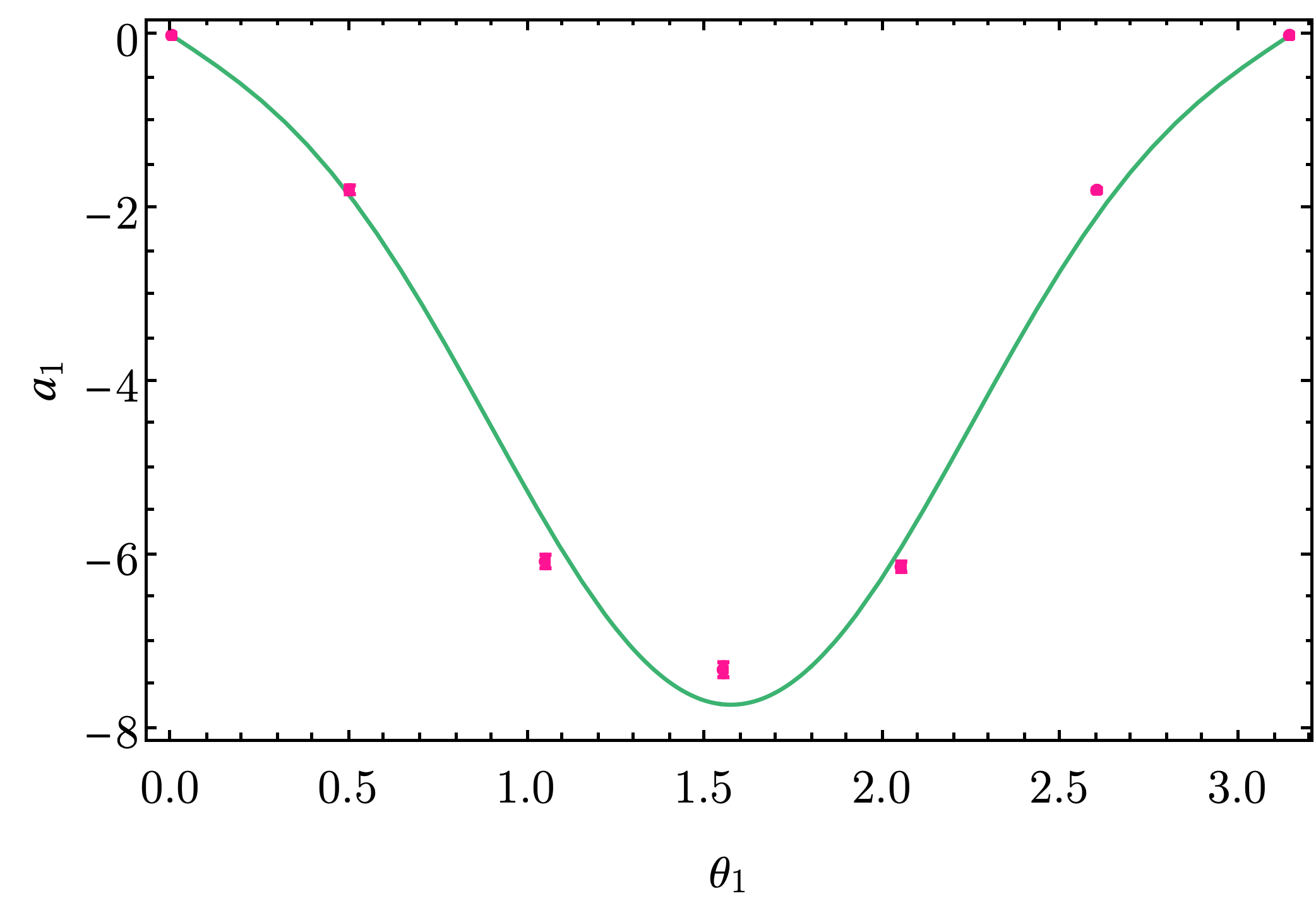}
    \caption{{Functional dependence of the $\eta^2 \chi_1^2$ term on $\theta_1$ (following the ansatz given by Eq.~\eqref{eq:fitachi}), as depicted in Eq.~\eqref{eq:deltasqfit}. Similar results are obtained for the remaining term of the parameterized fit for $\overline{\delta^2}$. All fits were performed with the \texttt{Fit} function in Mathematica \cite{fit}.
    }
    \label{fig:chi_coeffs}}
\end{figure}

The fitting procedure then follows the same structure described in the previous subsection. 
We first compute $\delta^2$ from Eq.~\eqref{eq:deltasq} for all the single spin simulations in our precessing dataset. We again neglect the in-plane spin angle $\phi_1$, reducing our independent variables to $(q, \chi_1,\theta_1)$. 
We then show $\overline{\delta^2}$ for fixed values of $\theta_1$, chosen to be the same values as for the energy: $\theta_1 \in \left\{\frac{\pi }{6},\frac{\pi }{3},\frac{\pi }{2},\frac{2 \pi }{3},\frac{5 \pi }{6}\right\}$. Again, at $\theta_1 \in \{0,\pi\}$, $\delta^2$ is defined to vanish so one recovers the non-precessing limit. 
In order to find an appropriate ansatz in this case, we start our procedure with $a\leq 3$ and $b\leq 2$. We then followed the iterative procedure described above to reduce our grid from 11 to 7 coefficients, representing an improvement of -16.5 in BIC. Then, for every fixed value of $\theta_1$, we have 
\begin{equation}
     \{a_i\}_{i=1}^{i=7}\left(
     \chi_1^2,\eta  \chi_1, \eta  \chi_1^2,\eta ^2 \chi_1, \eta^2 \chi_1^2,\eta ^3 \chi_1 ,\eta ^3 \chi_1^2
     \right).
    \label{eq:fitchi1}
\end{equation}
Finally, five more coefficients can be discarded when inspecting their $\theta_1$-dependence, resulting in a very simple ansatz for $\overline{\delta^2}$:
\begin{equation}
     \{a_i\}_{i=1}^{i=2}\left(\eta ^2 \chi_1 ^2,\eta ^3 \chi_1 ^2\right).
    \label{eq:fitchi}
\end{equation}
We propose again a sinusoidal ansatz for the coefficients $a_i$ which satisfies the boundary conditions:
\begin{equation}
     a_i(\theta_1)=A_i \sin \theta_1 + B_i \sin 3 \theta_1.
    \label{eq:fitachi}
\end{equation}
Figure~\ref{fig:chi_coeffs} shows an example of the functional dependence of the first fit coefficient $a_1$ ($\eta^3 \chi_1^2$) in terms of $\theta_1$. 

\begin{figure}   
    \includegraphics[width=1\columnwidth]{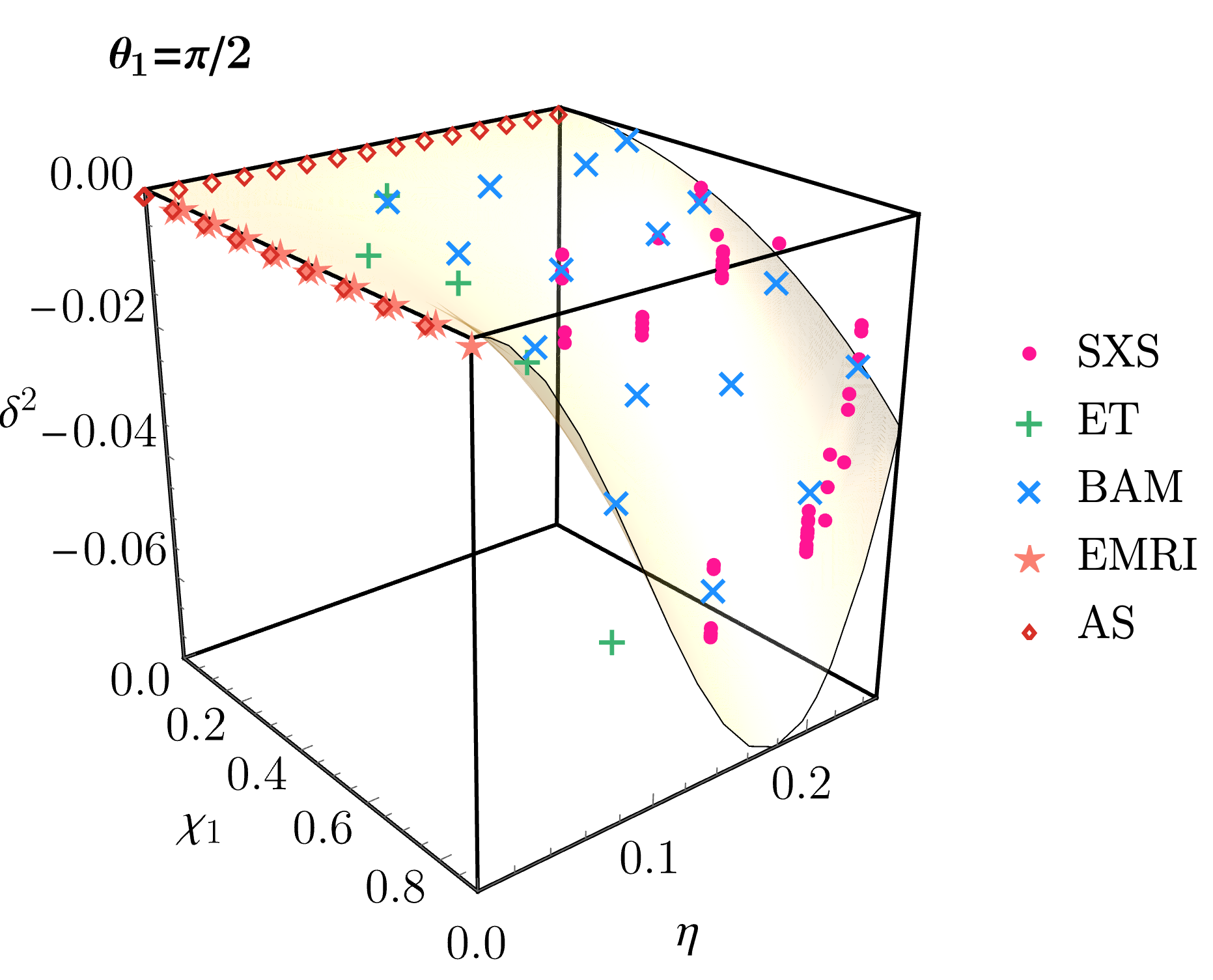}
    \caption{{
    Numerical evaluation of the parameterized fit of $\delta^2$ as defined in Eq.~\eqref{eq:deltasq}, obtained from the parameterized fit $\overline{\delta^2}$~\eqref{eq:deltasqfit} and $\delta^2_{\mathrm{EMR}}$ at a fixed spin orientation $\theta_1=\pi/2$, while varying the mass ratio $\eta$ and the spin magnitude $\chi_1$. The figure includes the single spin precessing simulations shown in Fig.~\ref{fig:datasetSS} that fall into this subspace.}
    \label{fig:chi_data_full}}
\end{figure}

The resulting fit is finally given by 
\begin{equation}
\begin{split}
    \overline{\delta^2}(\eta,\chi_1,\theta_1) =\ 
    &\eta ^2 \chi_1^2 \left[1.25552 \sin (3 \theta_1 )-6.47984 \sin (\theta_1 )\right]+\\
    &\eta ^3 \chi_1^2 \left[27.0456 \sin (\theta_1 )-4.71194 \sin (3 \theta_1 )\right],
    \label{eq:deltasqfit}
\end{split}
\end{equation}
and it is represented in Fig.~\ref{fig:chi_data_full} for $\theta_1=\pi/2$, together with the numerical $\delta^2$-values for the entire dataset. Note that the plot shows $\delta^2$, which is computed from Eq.~\eqref{eq:deltaEMRIfin}.
This fit can now be inserted into Eq.~\eqref{eq:deltasq} in order to get the new model for the remnant spin. 
In this expression, the aligned spin final spin dependence goes as $\chi_f^{\mathrm{AS}}\left(\eta,\chi_1\cos(\theta_1)\right)$ and the final mass is computed using the fit for $M_f(\eta,\chi_1,\theta_1)$ shown in Eq.~\eqref{eq:MffromErad}.

We can now assess the accuracy of our new model computing the final spin for our dataset using our new model, denoted as \texttt{PhenNew}. We then compare it with the current \phX model, which ignores the $\delta^2$ correction (\texttt{PhenXP}) and the \texttt{NRSur7dq4EmriRemnant} model.
\begin{figure}
    \includegraphics[width=1\columnwidth]{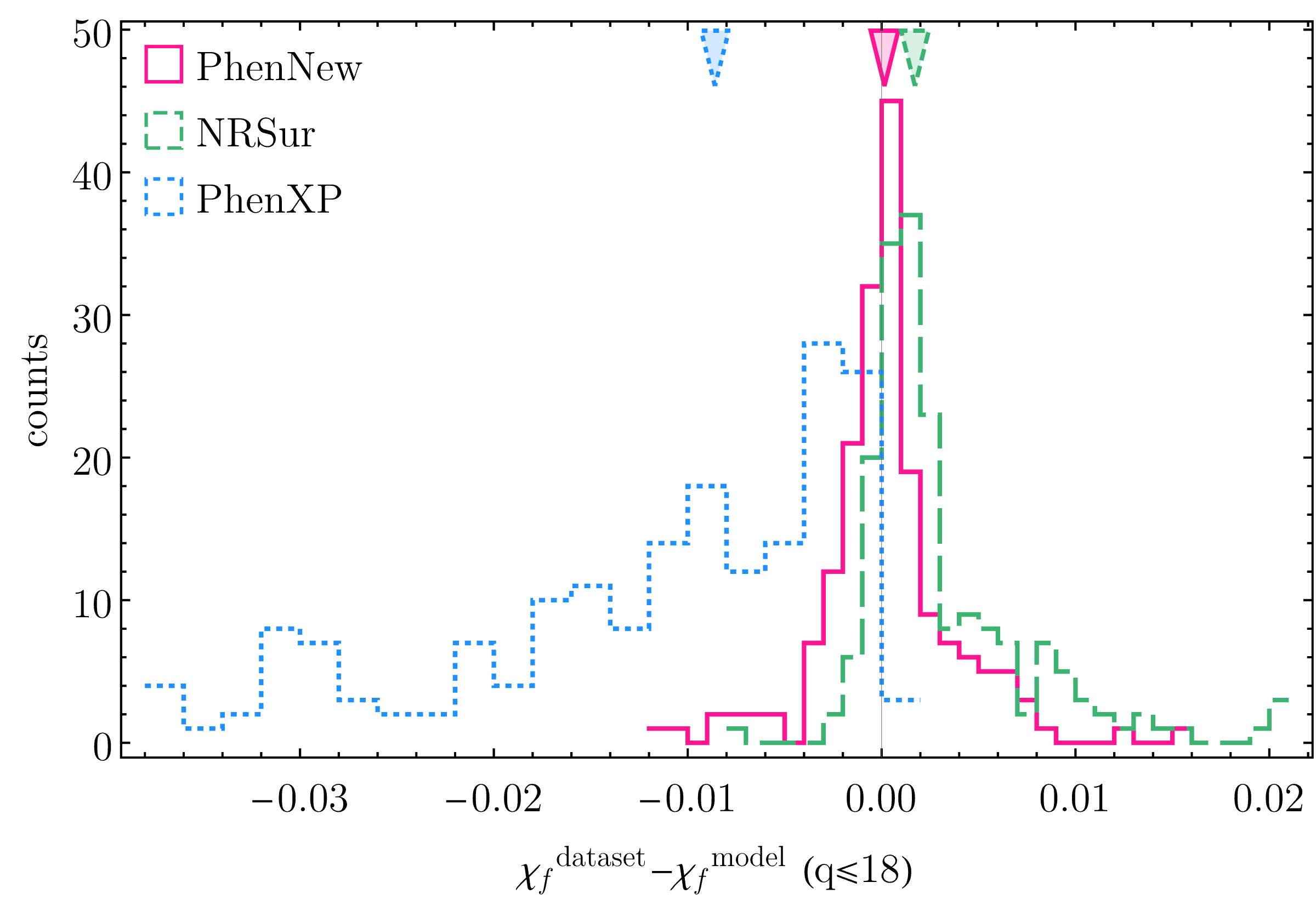}
    \caption{
    Histogram of the errors in the remnant spin computed with each of the three models for our single spin precessing dataset presented in Fig.~\ref{fig:datasetSS}. 
    The model developed in this project is labeled as \texttt{PhenNew}, the underlying model as \texttt{PhenXP} and \NRSurE as \texttt{NRSur}. 
    The triangles represent the median of each distribution, also included in Table~\ref{table:errorsmodel}. This table provides additional information on the distributions: the computational time needed to evaluate the dataset for each model, along with the root mean square errors (RMSEs).
    \label{fig:errorschi}}
\end{figure}
Figure~\ref{fig:errorschi} displays the error histogram for each model relative to the numerical values, equivalent to the approach in Fig.~\ref{fig:errorsE}, utilizing the full dataset without the EMRI waveforms. 
Table~\ref{table:errorsmodel} provides the median and root mean square errors, computed as in Eq.~\eqref{eq:RMSE}, of the distributions, along with the computational time required for each model evaluation. 
The \texttt{PhenXP} model for the final spin $\chi_f$ involves evaluating Eq.~\eqref{eq:deltasq} assuming $\delta^2=0$. On the other hand, \texttt{PhenNew} evaluates Eq.~\eqref{eq:deltasq} and the parameterized fit for $\overline{\delta^2}$~\eqref{eq:deltasqfit}, as well as $\delta^2_{\mathrm{EMR}}$, using the \texttt{KerrGeodesics} package. 
Both models are fully assessed in Mathematica, with most of the computational time attributed to solving the precessing geodesic equations to obtain $\delta^2_{\mathrm{EMR}}$.
In the case of \texttt{NRSur}, as previously mentioned, we evaluated the \NRSurE using the \texttt{SurfinBH} package, extracting the final mass and spin from the package's output.
Regarding computational times, it's important to note that the provided times are for the purpose of comparison and not aimed at optimizing the code's efficiency. In the case of the model developed in this project, the majority of the evaluation time is dedicated to solving the geodesic equations. As part of future work, we anticipate parameterizing $\Delta E_{\mathrm{EMR}}$ and $\delta^2_{\mathrm{EMR}}$ to make the fits entirely parametric.
The conclusions drawn from these results parallel those from the final mass: the new model offers a more accurate and less biased distribution than \phX, achieving precision comparable to that of the \NRSurE model while retaining the simplicity and efficiency of the model on which it is based.

To ensure completeness, we perform a final check on the extrapolation of this new remnant spin model for extreme spins (see App.~\ref{app:extrap}).
Once again, we confirm that our model is well-behaved even for extreme spins, despite not being calibrated in that regime, and it maintains the Kerr limit $|\chi_1|\leq 1$. 

\subsection{Cross-validation of the remnant model}\label{sec:cross}

In this last section we provide additional tests of our complete remnant model. Firstly, we compute "out-of-sample" errors to evaluate the consistency of our proposed ansätze. Finally, we test the performance of our model on the entire precessing dataset, including both double and single spin simulations.

The parameterized fits shown in Figs.~\ref{fig:E_full_data} and \ref{fig:chi_data_full} were obtained from our single spin precessing dataset (184 simulations), yielding the in-sample errors detailed in Table~\ref{table:errorsmodel}. 
To examine the consistency of our method, we now compute "out-of-sample" errors  using a procedure that involves dividing the data into 23 sets of 8 random samples each. For each set, we construct the final mass and spin magnitude fits using the remaining 176 data points and test their performance against the 8 validation samples.  
The resulting root mean square error values display the following mean values over the 23 sets: $\overline{\mathrm{RMSE}}(M_f)=6.3\cdot 10^{-4}$ and $\overline{\mathrm{RMSE}}(\chi_f)=3.2\cdot 10^{-3}$.
These values closely match those shown in Table~\ref{table:errorsmodel}, where no samples were taken to compute the fits. Therefore, we conclude that, as expected from our analysis based on information criteria, our models do not exhibit overfitting tendencies, affirming the consistency of our fitting procedure.

As a final test, we evaluate our remnant model for the complete precessing dataset outlined in Sec.~\ref{sec:NR}. 
Although the model has been calibrated for the single spin limit, one might be interested in its performance across the full precessing quasi-circular space, as well as its comparison with the currently available remnant models.
Figures~\ref{fig:errorsE} and \ref{fig:errorschi} reveal that \texttt{PhenXP} exhibits a biased distribution for both the final mass and spin magnitude, tending to overestimate the real value. 
The fitting quantities $\Delta E$ and $\delta^2$ consistently show a clear tendency in their sign, as evident in Figs.~\ref{fig:E_full_data} and \ref{fig:chi_data_full}.
These quantities appear as a small correction to the \texttt{PhenXP} model, which works for the double spin case, and hence the effect of our parameterized fits results in a slight shift towards the correct values. However, since they only account for the single spin, substantial improvements in accuracy are not expected in this scenario.
\begin{figure}   
    \includegraphics[width=1\columnwidth]{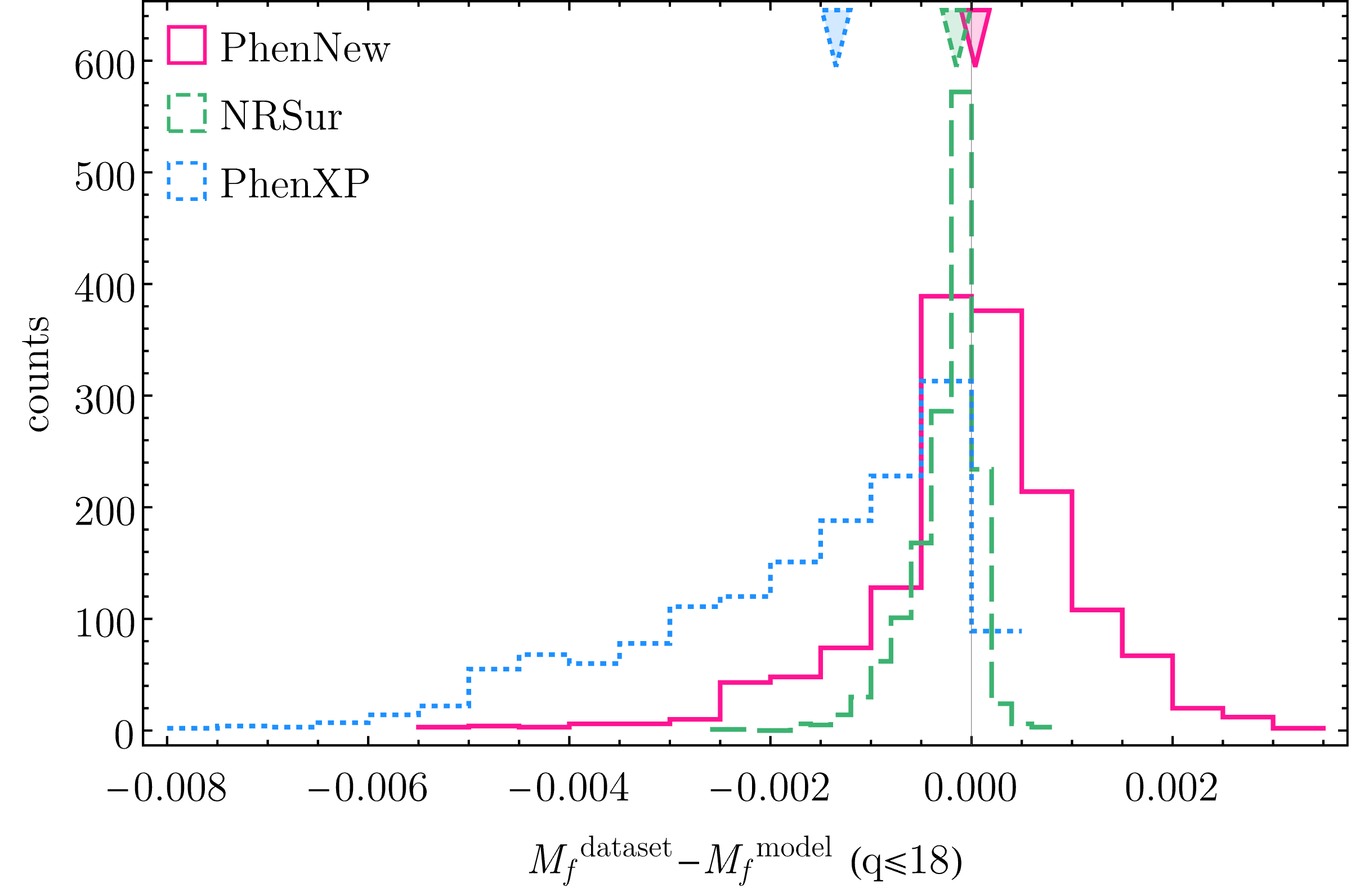}
    \includegraphics[width=1\columnwidth]{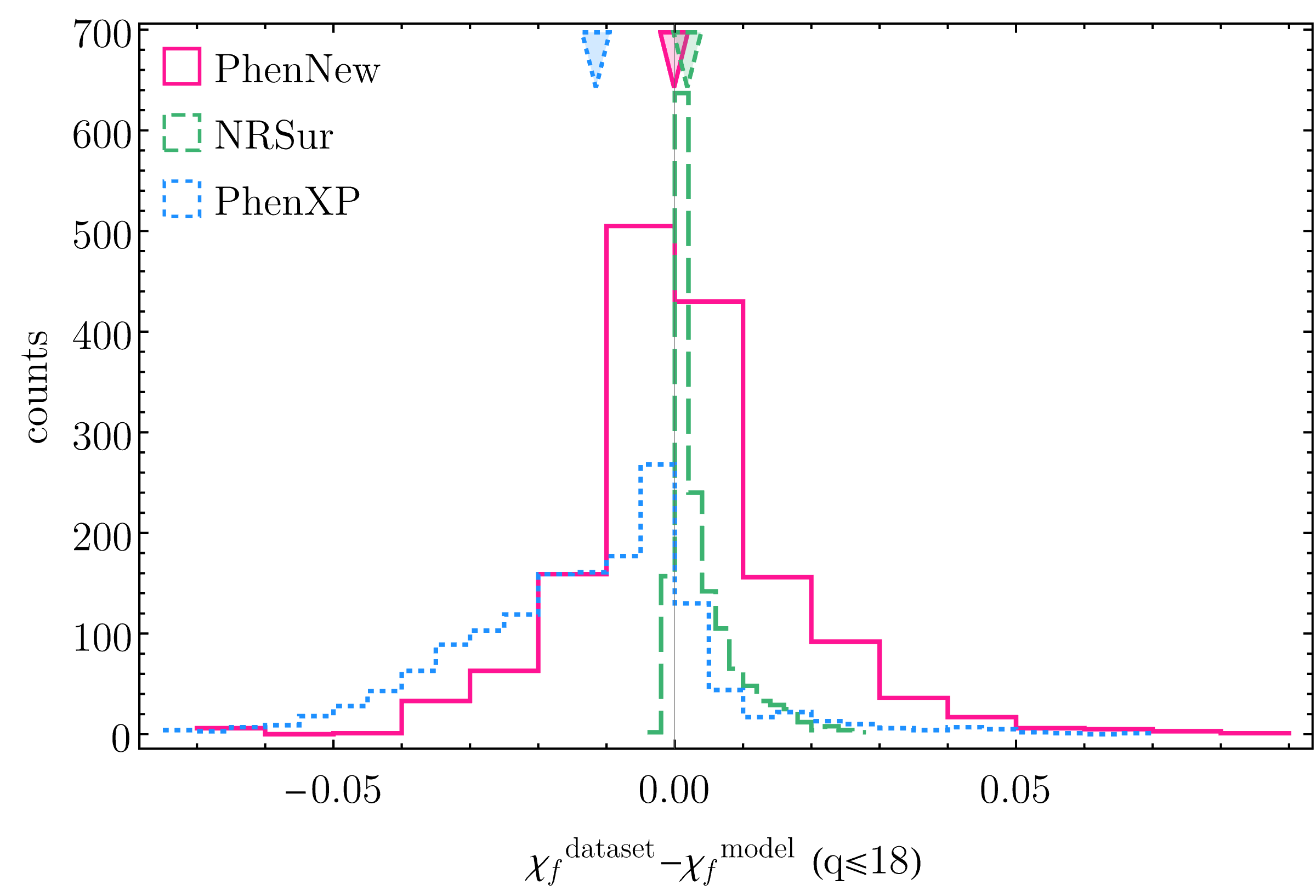}
    \caption{{Histograms of the errors in the remnant properties with each of the three models for our full precessing dataset presented in Fig.~\ref{fig:dataset}. \texttt{PhenNew} corresponds to \texttt{PhenXP} plus the corrections derived in Secs.~\ref{sec:finalmass} and \ref{sec:finalspin} for the final mass (top panel) and final spin (bottom), respectively, and \texttt{NRSur} corresponds to the \NRSurE model. 
    The triangles represent the median of each distribution. 
    Table~\ref{tab:errorsmodelDS} provides the medians along with the root mean square errors (RMSEs) for each distribution.}
    \label{fig:double_spin}}
\end{figure}
\begin{table}
\begin{tabular}{ccccc}
\cline{3-5}
\noalign{\vskip\doublerulesep
         \vskip-\arrayrulewidth}
\cline{3-5}       
 & & \texttt{PhenNew} & \texttt{PhenXP} & \texttt{NRSur} \\ \hline \hline
\multirow{2}{*}{$M_f$} & Median & $3.7\cdot 10^{-5}$ & $-1.3\cdot 10^{-3}$ & $-1.5\cdot 10^{-4}$ \\
& RMSE & $1.1\cdot 10^{-3}$ & $2.3\cdot 10^{-3}$ & $4.2\cdot 10^{-4}$ \\ \hline
\multirow{2}{*}{$\chi_f$} & Median & \, $-9.9\cdot 10^{-5}$ \, & \, $-1.2\cdot 10^{-2}$ \, & \, $1.8\cdot 10^{-3}$ \, \\
& RMSE & $1.6\cdot 10^{-2}$ & $2.2\cdot 10^{-2}$ & $6.1\cdot 10^{-3}$ \\ \hline \hline
\end{tabular}    
\caption{Median value and root mean square error (RMSE) of the error distributions of the remnant mass $M_f$ and spin magnitude $\chi_f$ for different models, with respect to the full numerical relativity dataset shown in Fig.~\ref{fig:dataset}. The histograms of the distributions are shown in Fig.~\ref{fig:double_spin}, top and lower panel, respectively.
\label{tab:errorsmodelDS}}
\end{table}

Figure~\ref{fig:double_spin} and Table~\ref{tab:errorsmodelDS} demonstrate that the error distributions obtained with \texttt{PhenNew} for the complete precessing dataset exhibit less bias compared to those obtained with \texttt{PhenXP}. 
However, given that \texttt{PhenNew} only considers the single spin correction, the overall performance does not exhibit a substantial improvement, as anticipated. Incorporating the double spin correction remains a direction for future work.
In Fig.~\ref{fig:double_spin}, we observed a slightly biased error distribution of the remnant properties for \NRSurE. Despite extensive tests, including waveform alignment using quadrupole alignment and different rotation methods, the small magnitude of the bias (see Table~\ref{tab:errorsmodelDS}) makes it challenging to track down the source of the error.
Moreover, Ref.~\cite{NRSur7dq4EmriRemnant} only provides absolute errors, which limits further comparisons on the relative error distributions found.
While it would be interesting to further understand this issue, it is not within the scope of this project, and further exploration remains a topic for future work.

\section{Conclusions}
\label{sec:conclusions}
%%%%%%%%%%%%%%%%%%%%%%%

In this work, we presented a new parameterized remnant model for single spin precessing black hole binary systems at any mass ratio. An efficient and accurate remnant model for precessing systems is a key component for the advancement of precessing waveform models. Specifically, given that the ringdown frequencies are entirely characterized by the final state of the binary, remnant models play a vital role in conducting tests of general relativity through ringdown studies.

We employed information from precessing geodesics at the ISCO to estimate the remnant properties in the extreme mass ratio regime, and numerical data at mass ratio 1000 as a cross-check. For the final mass we have fixed the term linear in symmetric mass ratio $\eta$ to the values obtained from the EMR limit, and for the final spin we have fixed both the linear and quadratic terms in $\eta$ terms this way.
As has become customary before, we have worked  in a co-orbital frame, which drastically reduces differences between the aligned spin and precessing sectors, and in addition we have subtracted previous aligned spin fits from our data before performing the fit to the precessing dataset. These procedures have allowed us to obtain rather accurate but simple fits from a relatively small number of numerical relativity waveforms across the entire range of mass ratios. Overfitting was controlled by model selection based on the BIC information criterion, and cross-checked by performing ``out-of-sample'' errors  tests which confirm the consistency of our remnant models, yielding RMSEs very close to those obtained for the model utilizing the full dataset.

We observe that the correction terms that map aligned spin results to the precessing case have a dominant sign. In consequence, our models remove biases that had been present in previous simple fits that only used aligned spin numerical relativity data and are being used in precessing phenomenological waveform models ~\cite{IMRPhenomXPHM,IMRPhenomTPHM}.
These biases have found to be related to the typical positive sign in $\Delta E$ and negative in $\delta^2$, leading to an overestimation trend in the underlying model. 

The assessment of the resulting models is summarized in Table~\ref{table:errorsmodel}. Our new model \texttt{PhenNew} surpasses its underlying baseline \texttt{PhenXP} used in current phenomenological models, achieving a performance akin than \NRSurE (\texttt{NRSur}) at much smaller computational cost. The largest contribution to the computational cost is the evaluation of the final mass and spin contributions of the geodesic approximation. While our straightforward Mathematica code could be optimized, or implemented in other languages, a further path to accelerating the evaluation would be to make a fast approximate model of the geodesic contribution. 
The evaluation of our fits on the complete precessing dataset reveals that the new model, \texttt{PhenNew}, exhibits a less biased distribution compared to \texttt{PhenXP}, but, not surprisingly, lacks a significant overall improvement for the double spin case. In order to develop a general model of precessing remnant mass and spin, the next steps will therefore be to extend  our work to the double spin case, and also to include the dependency on the in-plane spin angles.

In Sec.~\ref{sec:construction} we have discussed the problem of creating a consistent precessing dataset from several different numerical relativity catalogs. Here we have only used this heterogeneous dataset for models of the final state, using a reference time of $100 M$ before the merger to define the input data for our fits to facilitate comparisons with \NRSurE \cite{NRSur7dq4EmriRemnant}. Future work should investigate the optimization of trade-offs related to the choice of reference time: later times minimize the spin dynamics between input parameters and results, whereas earlier times benefit the connection of the final state fits with models for the inspiral. Furthermore, ambiguities arise in the definition of the merger time across different datasets, and in the choice of co-orbital frame. Future work will investigate these issues further.
We have also created consistent datasets for the waveforms, which we have not discussed and used in this paper. This  dataset is intended for the calibration of precessing waveform models to numerical relativity, where a large number of waveforms is required, and the pooling of data from different numerical relativity catalogues will be advantageous.

\section*{Acknowledgements}
We would like to thank Scott A. Hughes, Anuj Apte, Gaurav Khanna and Halston Lim for providing the EMRI waveforms used in this project; Maite Mateu-Lucena for being involved in early stages of producing the dataset; Isabel Suárez-Fernández for running some of the Einstein Toolkit simulations; and Anna Heffernan, Antoni Ramos-Buades, Cecilio García-Quirós and Vijay Varma for useful comments and discussions. This work makes use of the Black Hole Perturbation Toolkit~\cite{BHPToolkit}, in concrete the \hyperlink{https://zenodo.org/records/8108265}{\texttt{KerrGeodesics}} package.
The authors thankfully acknowledge the computer resources at MareNostrum and the technical support provided by Barcelona Supercomputing Center (BSC)  through funding from the Red Española de Supercomputación (RES).
Maria de Lluc Planas is supported by the Spanish Ministry of Universities via an FPU doctoral grant (FPU20/05577).
Joan Llobera-Querol is supported by the Comunitat Autònoma de les Illes Balears through the Direcció General de Recerca, Innovació i Transformació Digital via an FPU doctoral grant FPI/2022.
This work was supported by the Universitat de les Illes Balears (UIB); the Spanish Agencia Estatal de Investigación grants PID2022-138626NB-I00, PID2019-106416GB-I00, RED2022-134204-E, RED2022-134411-T, funded by MCIN/AEI/10.13039/501100011033; the MCIN with funding from the European Union NextGenerationEU/PRTR (PRTR-C17.I1); Comunitat Autonòma de les Illes Balears through the Direcció General de Recerca, Innovació I Transformació Digital with funds from the Tourist Stay Tax Law (PDR2020/11 - ITS2017-006), the Conselleria d’Economia, Hisenda i Innovació grant numbers SINCO2022/18146 and SINCO2022/6719, co-financed by the European Union and FEDER Operational Program 2021-2027 of the Balearic Islands; the “ERDF A way of making Europe”.
This material is based upon work supported by NSF's LIGO Laboratory which is a major facility fully funded by the National Science Foundation.

\appendix

\section{Details on the general geodesic equations}
\label{app:geodesics}
%%%%%%%%%%%%%%%%%%%%%%%%%%%%%%%%%%%%%
Following Ref.~\cite{Schmidt_2002}, for geodesics in Kerr spacetime the constants of motion $E$, $L_\mathrm{z}$ and Carter's constant $Q$ for given orbital parameters and also the parameters of the source $a$ and $q=m_1/m_2=M/\mu$ are given by (using Boyer-Lindquist coordinates $(r,\theta,\phi, t)$)
\begin{equation}
    \frac{\mathrm{d}r}{\mathrm{d}\tau}=R(r)=[(r^2+a^2)E-aL_{\mathrm{z}}]^2-\Delta [\mu^2r^2+(L_{\mathrm{z}}-aE)^2+Q]=0,
    \label{eq:R=0}
\end{equation}
\begin{equation}
    \frac{\mathrm{d}\theta}{\mathrm{d}\tau}=\Theta(\theta)=Q-\left[(\mu^2-E^2)a^2+\frac{L_{\mathrm{z}}^2}{\sin^2\theta}\right]\cos^2\theta=0,
    \label{eq:theta=0}
\end{equation}
where $\Delta=r^2-2Mr+a^2$.
The roots of the equations correspond to the turning points of the radial and polar motion. For circular orbits ($e=0$), we will need a third constraint $R'(r_0)=0$, apart from $R(r_0)=0$ and $\Theta (\theta_0)=0$. 
Circular orbits are stable if $R''(r_0)<0$. The radius which separates the stable and unstable orbits is known as the ISCO and hence satisfies $R''(r_0)=0$. Thus, solving these four equations one can find the radius $r_0=r_{\mathrm{ISCO}}$, the energy $E$, the orbital angular momentum $L_{\mathrm{z}}$ and Carter's constant $Q$ for a given system and orbital quantities.

If we use the dimensionless quantities introduced in Eq.~\eqref{eq:dimless} and rearrange $\Theta(\theta_{-})=0$, we can express the Carter's constant as
\begin{equation}
    \tilde{Q}=\cos^2\theta_{-}\left[\tilde{a}^2(1-\tilde{E}^2)+\frac{\tilde{L_{\mathrm{z}}^2}}{1-\cos^2 \theta_{-}}\right].
\end{equation}
Substituting this equation in $\tilde{R}(\tilde{r})$ gives
\begin{equation}
    \tilde{R}(\tilde{r})=f(\tilde{r})\tilde{E}^2-2g(\tilde{r})\tilde{E}\tilde{L_{\mathrm{z}}}-h(\tilde{r})\tilde{L_{\mathrm{z}}}+d(\tilde{r}),
    \label{eq:R(r)}
\end{equation}
where
\begin{align}
f(\tilde{r})&=\tilde{r}^4+\tilde{a}^2[\tilde{r}(\tilde{r}+2)+\cos^2\theta \tilde{\Delta}], \\
    g(\tilde{r})&=2\tilde{a}\tilde{r},\\
    h(\tilde{r})&=\tilde{r}(\tilde{r}-2)+\frac{\cos^2\theta_-}{1-\cos^2\theta_-}\tilde{\Delta},\\
    d(\tilde{r})&=(\tilde{r}^2+\tilde{a}^2 \cos^2\theta_-)\tilde{\Delta},
\end{align}
and one can then compute $\tilde{R}'(\tilde{r_0})$ and $\tilde{R}''(\tilde{r_0})$ from Eq.~\eqref{eq:R(r)}.
Solving these equations yields to \emph{four} solutions for the constants of motions, and one fixed value for the $r_{\mathrm{ISCO}}$ in the case of circular orbits. Considering only those solutions with positive energy, we have $(\tilde{E}^{(p)},\tilde{L_{\mathrm{z}}}^{(p)},\tilde{Q}^{(p)})$ and $(\tilde{E}^{(r)},\tilde{L_{\mathrm{z}}}^{(r)},\tilde{Q}^{(r)})$, where $p$ stands for prograde orbits and $r$, for retrograde. It is verified that $\tilde{E}^{(p)}<\tilde{E}^{(r)}$ and $\tilde{L_{\mathrm{z}}}^{(p)}<\tilde{L_{\mathrm{z}}}^{(r)}$- for prograde orbits the particle has higher binding energy and co-revolves with the black hole, whereas retrograde orbits usually counter-revolves.

\section{Information criteria}\label{ap:IC}
%%%%%%%%%%%%%%%%%%%%%%%%%%%%%%%%%%%%%%%%%%%

We do not work with an a-priori ansatz for our parameterized fits, but rather select the best functional form from a wide class of models. We then use information criteria to perform the model selection to avoid overfitting, following \cite{Jimenez-Forteza:2016oae}, where aligned spin fits for the remnant quantities are constructed. In this appendix we describe the information criteria in more detail.

A basic performance metric for model adjustment is the root mean square error (RMSE). For a model of a quantity $q$ dependent on parameters $\lambda$, and data points $\left(\lambda_i, q_i\right)$ for $i~=~1\dots N$,
\begin{equation}\label{eq:RMSE}
    \mathrm{RMSE}\left[\mathrm{model}\right] = \sqrt{ \dfrac{1}{N} \sum_{i=1}^{N} \left[ q_i - \mathrm{model}(\lambda_i) \right]^2 }.
\end{equation}
Using only the RMSE to perform model selection is prone to overfitting. 
For this reason, when selecting the best model, one should penalize models according to its complexity, specifically the number of free coefficients. A widely-used statistical quantity is the Akaike Information Criterion (AIC) \cite{AIC},
\begin{equation}
    \mathrm{AIC} = -2\ln{\mathcal{L}_{max}} + 2N_\mathrm{coeffs},
\end{equation}
which compensates the accuracy of the fit with the number of coefficients.
Here a lower value of the AIC indicates better suitability of the model. We have used the implementation of the AIC in the Mathematica's \texttt{LinearModelFit} function ~\cite{linearmodelfit}.

An alternative quantity that serves the same purpose, but has a different theoretical foundation, is the Bayesian Information Criterion (BIC) \cite{BIC}
\begin{equation}
    \mathrm{BIC} = -2\ln{\mathcal{L}_{max}} + N_\mathrm{coeffs} \ln{N_\mathrm{data}}.
\end{equation}
In terms of performance, both criteria penalize the degeneracies between parameters and the BIC usually penalizes additional parameters more than AIC.
To discern between models, a 1 unit difference is generally required, while 10 points would be decisive evidence. Further discussion on the criteria can be found in \cite{Criteria}.

\section{Fit for the updated aligned radiated energy}\label{ap:fits}
%%%%%%%%%%%%%%%%%%%%%%%%%%%%%%%

In this appendix we provide the full parameterized expression for the aligned spin radiated energy used in this paper, which updates the one developed in Ref.~\cite{Jimenez-Forteza:2016oae}.

Defining
\begin{equation}
    \Shat := \dfrac{\cone+q^2\ctwo}{1+q^2},
\end{equation}
we can write:

\begin{widetext}
\begin{multline}
    E_{\mathrm{AS}} = 0.288265 \eta^5 (\cone-\ctwo)^2 - 0.0483974 \eta^2 (1-1.76539\eta) \sqrt{1-4\eta} (\cone-\ctwo) + \\
    +\ \dfrac{1393.61 \eta^7 - 1160.94 \eta^6 + 372.473 \eta^5 - 54.0578 \eta^4 + 3.33345 \eta^3 + 0.44487 \eta^2 + \left(1-\frac{2\sqrt{2}}{3}\right) \eta}{\left(1.96359 \eta ^2+0.557424 \eta -0.956935\right) \Shat +1} \\
    \left[ \left(-5.40979 \eta^2 + 1.74325 \eta - 0.106587\right) \Shat^6 + \left(0.915964 \eta ^2+0.0338535 \eta -0.0809724\right) \Shat^5 + \right. \\
    \left. + \left(3.93186 \eta ^2-1.15351 \eta +0.0316422\right) \Shat^4 + \left(-1.16612 \eta ^2+0.379967 \eta -0.0552524\right) \Shat^3 + \right. \\
    \left. + \left(-0.950876 \eta ^2+0.635553 \eta -0.173169\right) \Shat^2 + \left(-2.75115 \eta ^2+1.73637 \eta -0.398234\right) \Shat + 1 \right].
    \label{eq:EradASfit}
\end{multline}
\end{widetext}
\section{Extrapolation of the remnant model towards extreme spins}
\label{app:extrap}
In this appendix we assess the extrapolation behaviour of our parameterized remnant models focusing on the scenario where precession effects are maximized. 
Specifically, we explore spin magnitudes beyond the calibrated regime ($0.8\leq \chi_1 \leq 1$), considering the case of in-plane configurations ($\theta_1=\pi/2$).

\begin{figure*} 
    \centering
    \includegraphics[width=0.82\columnwidth]{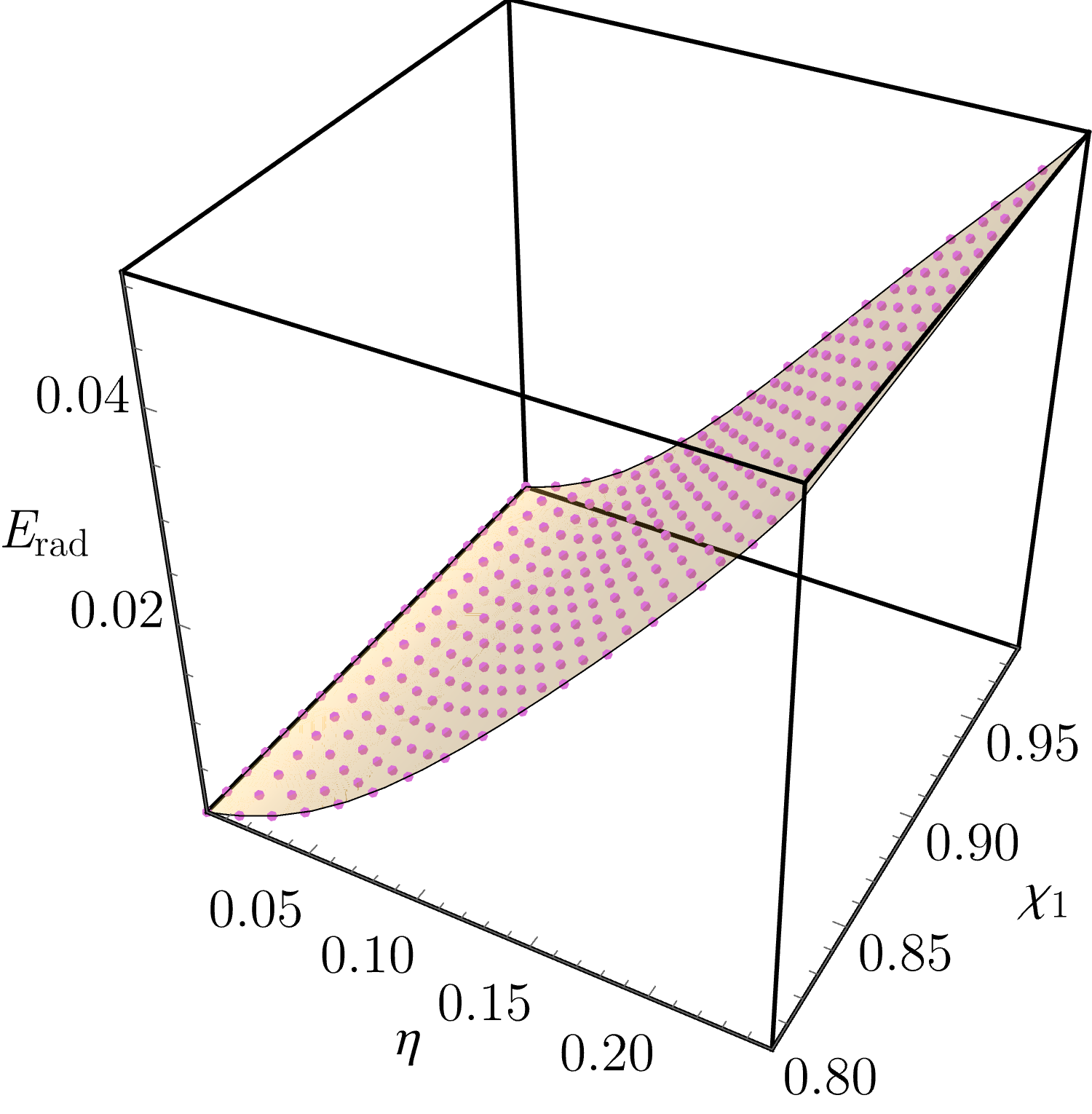}
    \hspace{2.5cm}
    \includegraphics[width=0.8\columnwidth]{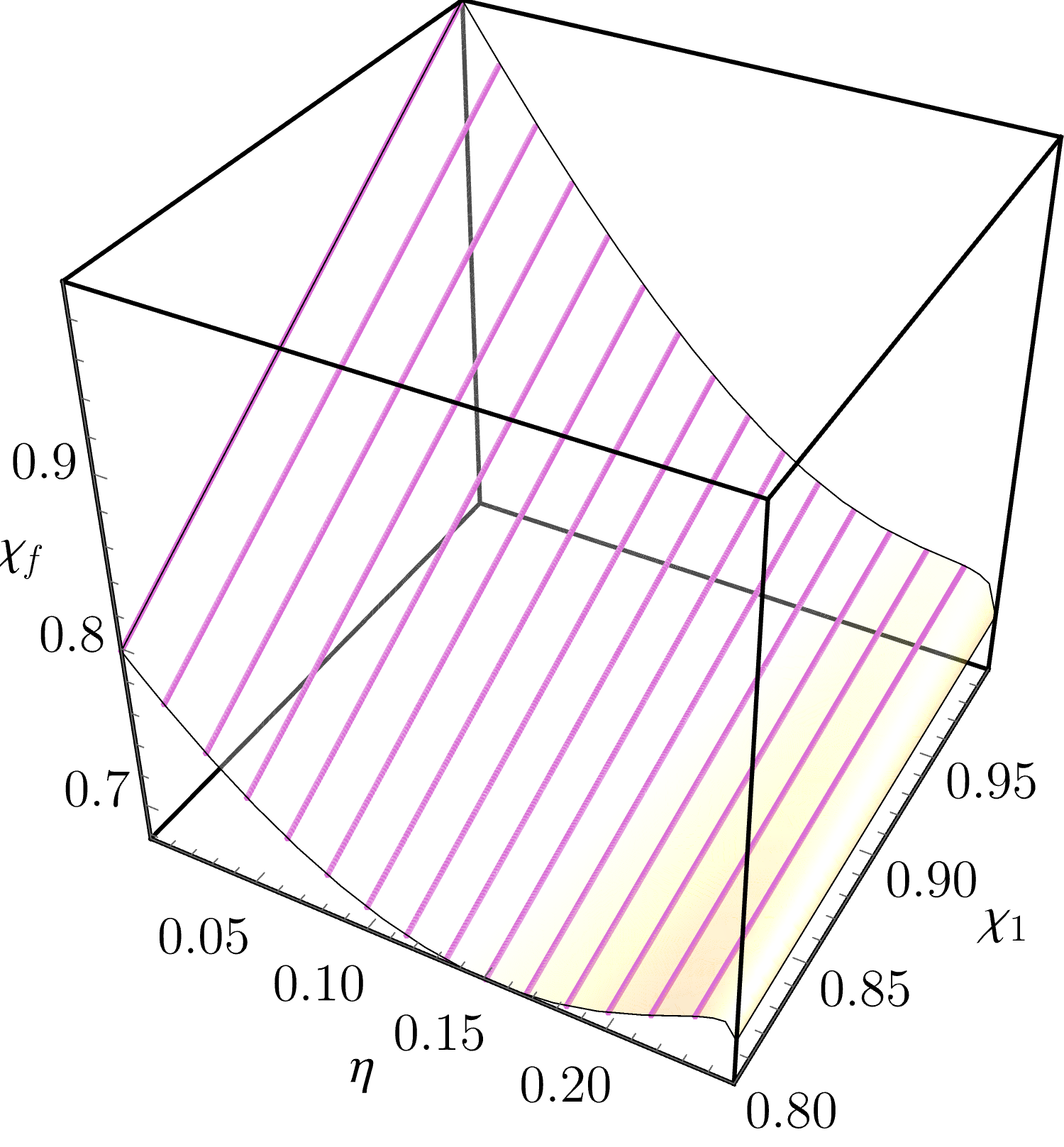}
    \caption{{Extrapolation of the remnant fits developed in Secs.~\ref{sec:finalmass} (left panel) and \ref{sec:finalspin} (right panel) outside of their spin calibration regime, $0.8\leq \chi_1 \leq 1$, for a single in-plane spin configuration ($\theta_1=\pi/2$), where precession effects are maximized.}
    \label{fig:extrap}}
\end{figure*}

Figure~\ref{fig:extrap} illustrates the extrapolation results, showing a smooth continuation without exhibiting any nonphysical behaviour, as well as the Kerr limit $\chi_1\leq 1$.
While these extrapolations provide valuable insights, it is important to interpret them cautiously, recognizing the need for further refinement when numerical data becomes available in the high spin magnitude regime.

% ~~~~~~~~~~ References ~~~~~~~~~~ %

% Try this if the last two columns before the bib are not breaking nicely.
%\vspace{0.1in}

%\vfil

\let\c\Originalcdefinition %
\let\d\Originalddefinition %
\let\i\Originalidefinition

\bibliography{bibliography}
% ~~~~~~~~~~ END DOCUMENT ~~~~~~~~~~ %

\end{document}

%% file: UsePackages.tex
\usepackage[T3,T1]{fontenc}
\usepackage{mathrsfs} % For \mathscr fonts
\usepackage{bm} % For \bm fonts
\makeatletter
\@ifclassloaded{beamer}
  { % if this is a beamer document, use sans-serif fonts
    \typeout{UsePackages: Detected beamer}
    \usepackage{tgheros}
    
  }
  { % otherwise:
    \typeout{UsePackages: Did not detect beamer}
    \ifx\asybeamer\undefined % use times for articles
    \typeout{UsePackages: Detected article}
    \usepackage[varg]{txfonts} % For math
    \usepackage{tgtermes} % For text
    \else % use beamer fonts for asy documents with beamer
    \typeout{Fonts: Detected asy for beamer}
    \usepackage{tgheros}
    
    \fi
  }
\usepackage{microtype} % For nicer alignment of text
\def\MT@register@subst@font{
  \MT@exp@one@n\MT@in@clist\font@name\MT@font@list
  \ifMT@inlist@\else\xdef\MT@font@list{\MT@font@list\font@name,}\fi}
\makeatother

\usepackage{graphicx} %
\usepackage[x11names,svgnames,rgb]{xcolor} %
\usepackage{xspace}
\usepackage{braket}
\usepackage{accents}
\usepackage{siunitx}
\usepackage{mathtools}
\DeclareSymbolFontAlphabet{\mathrm}{operators}

\definecolor{CiteColor}{rgb}{0.18039, 0.18824, 0.57255}
\definecolor{UrlColor} {rgb}{0.741, 0.173, 0.000}
\definecolor{DarkUrlColor} {rgb}{0.500, 0.110, 0.000}
\definecolor{LinkColor}{rgb}{0.25098, 0.47843, 0.04706}
\makeatletter %
\newcommand{\ShowFont}{%
  \typeout{The main font is \f@encoding \space \f@family \space %
    \f@series \space \f@shape \space at \f@size pt.}%
  \typeout{The math font sizes are \tf@size pt (main), \sf@size pt %
    (script), and \ssf@size pt (scriptscript).}%
  \typeout{The linewidth is \the\linewidth}} %
\makeatother %
\usepackage{xargs}